\documentclass[
 reprint,
 amsmath,amssymb,
 aps,
 prx,
 superscriptaddress
]{revtex4-2}
\usepackage{float}
\usepackage{CJKutf8}
\usepackage[normalem]{ulem}
\usepackage[english]{babel}
\usepackage{microtype}   
\usepackage[utf8]{inputenc}
\usepackage{amsmath, amssymb, amsfonts}
\usepackage{physics}         
\usepackage{bm}              
\usepackage{graphicx}
\usepackage[normalem]{ulem}
\usepackage{hyperref}
\usepackage[percent]{overpic}
\AtBeginDocument{
  \setlength{\abovedisplayskip}{6pt}
  \setlength{\belowdisplayskip}{6pt}
  \setlength{\abovedisplayshortskip}{4pt}
  \setlength{\belowdisplayshortskip}{4pt}
}

\usepackage{indentfirst}
\usepackage[dvipsnames]{xcolor}
\usepackage[T1]{fontenc}
\definecolor{darkblue}{RGB}{0,0,149}
\usepackage{subcaption}
\usepackage{caption}
\usepackage{ragged2e}
\DeclareCaptionJustification{justified}{\raggedright}
\hypersetup{
    colorlinks=true,
    linkcolor=red,
    filecolor=darkblue,
    urlcolor=darkblue,
    citecolor=Green,
    linktocpage=true,
}
\usepackage{tikz}
\usetikzlibrary{decorations.pathreplacing}
\usepackage{adjustbox}
\usepackage{url}

\begin{document}
\raggedbottom                 
\setlength{\parskip}{0pt}     
\setlength{\parindent}{15pt}  

\title{Coherence-Enhanced Quantum Battery Charging with Ergotropy Stabilization}

\author{Fan Yang}
\affiliation{Institute for Quantum Science and Engineering, Texas A\&M University, College Station, Texas 77843, USA}
\author{Hui Wang }
\email{huiwangph@gmail.com}
\affiliation{Institute for Quantum Science and Engineering, Texas A\&M University, College Station, Texas 77843, USA}
\author{Yusef Maleki}
\affiliation{Institute for Quantum Science and Engineering, Texas A\&M University, College Station, Texas 77843, USA}
\author{William J. Munro}
\affiliation{Okinawa Institute of Science and Technology Graduate University, Onna-son$,$ Okinawa 904-0495$,$ Japan}
\author{Girish S. Agarwal}
\affiliation{Institute for Quantum Science and Engineering, Department of Physics and Astronomy, Department of Biological and Agricultural Engineering, Texas A\&M University, College Station, Texas 77843, USA}
\author{Marlan O. Scully}
\affiliation{Institute for Quantum Science and Engineering, Texas A\&M University, College Station, Texas 77843, USA}
\affiliation{Baylor University, Waco, TX 76798, USA}
\affiliation{Princeton University, Princeton, New Jersey 08544, USA}

\date{\today}

\begin{abstract}
Quantum batteries utilize nonclassical resources to achieve charging speed and energy storage performances that surpass classical thermodynamic limits. However, the practical realization of quantum batteries is often constrained by the inevitable environment-induced dissipation of both stored ergotropy and coherence. To actively counteract these losses, we propose a dual-channel coherence framework that exploits dark-state protection to stabilize ergotropy. We conduct, for the first time, an investigation of the synergistic interplay between internal charger coherence and reservoir squeezing, the latter acting as a source of external coherence. In the resource-efficient regime where charger and battery sizes are comparable, our study shows that internal charger coherence and reservoir squeezing jointly enhance the transient charging power. Crucially, initial charger coherence is the fundamental resource for maximizing and stabilizing steady-state ergotropy through dark-state protection. Our analysis reveals that these advantages are driven by the buildup of local battery coherence, which emerges from the integration of both internal and external coherence sources. These results offer a robust pathway for high-power, stabilized energy storage in quantum architectures.

\end{abstract}

\maketitle

\section{Introduction}


Quantum mechanical principles have revealed profound advantages across diverse technological frontiers. In the realm of energy storage and work extraction, groundbreaking advances in quantum thermodynamics have highlighted the essential role of nonclassical properties. Specifically, resources such as internal state coherence can fundamentally increase the work-extraction capacity of atomic quantum systems~\cite{doi:10.1126/science.1078955}. Optical squeezing, on the other hand, introduces quantum correlations and provides a powerful tool for reservoir engineering, acting as a source of external coherence for the atomic system~\cite{agarwal1990cooperative}. Another key resource is Dicke superradiance~\cite{dicke1954coherence}, based on the collective coupling of an ensemble of uncoupled two-level atomic systems to a common optical mode~\cite{ferraro2018high,campaioli2017enhancing}. 

Quantum batteries (QBs)~\cite{alicki2013entanglement,campaioli2024colloquium,ferraro2026opportunities} serve as a central paradigm for managing energy storage and the subsequent performance of work, utilizing such nonclassical resources to achieve performance surpassing classical limits. While the thermodynamic utility of internal quantum correlations like entanglement remains highly debated~\cite{andolina2019extractable,kamin2020entanglement,tabesh2020environment}, quantum coherence has emerged as a much more definitive resource, exhibiting a direct and robust relationship with enhanced battery efficiency and work-extraction capacity~\cite{ferreri2025quantum,kamin2020entanglement, centrone2023charging,li2025enhancing,lai2024quick}. Despite these advancements, a unified framework leveraging multiple coherence sources to enhance QB performance remains largely unexplored. In this work, we propose a framework in which coherence is supplied through two sources: reservoir squeezing as an external coherence source and coherent preparation of the charger's initial state as an internal one.
This dual-source approach, facilitated by dark-state protection (see below), simultaneously accelerates charging and enhances long-term ergotropy storage.

The operational performance of QBs is evaluated using ergotropy—the maximum work extractable from a quantum state via cyclic unitary operations \cite{allahverdyan2004maximal}. Total ergotropy fundamentally decomposes into two distinct components: one part arising from population inversion, and another originating from quantum coherence~\cite{francica2020quantum}. Coherence is essential for achieving distinct quantum advantages in energy storage and work extraction. However, both ergotropy components are inherently fragile in realistic open systems. Environmental coupling inevitably dissipates both the energetic population and phase coherence, rapidly depleting the stored ergotropy and severely restricting the QB's practical utility.

Mitigating the inevitable dissipation of stored ergotropy remains a central challenge for practical QBs. While several strategies have been proposed to preserve extractable work, they often impose severe operational trade-offs. For instance, encoding the battery within decoherence-free subspaces restricts its coupling to external fields, making efficient charging highly challenging \cite{liu2019loss}. Conversely, strategies based on sequential measurements \cite{gherardini2020stabilizing} or active feedback control require external energy costs that can reduce the net energy gain \cite{mitchison2021charging}. These limitations suggest that many stabilization protocols effectively force a trade-off between charging efficiency and storage longevity. One way to avoid these trade-offs is passive stabilization via dark-state protection in a shared reservoir \cite{pirmoradian2019aging,quach2020using,tabesh2020environment}, avoiding the continuous cost associated with the continuous-squeezing protocol~\cite{ferraro2018high,campaioli2017enhancing,gyhm2022quantum,binder2015quantacell}. 
Earlier dark-state quantum-battery studies generally required a charger significantly larger than the battery to obtain appreciable steady-state ergotropy. In contrast, we focus on the resource-efficient regime in which charger and battery sizes are comparable, motivated by the broader need to reduce energy and hardware overhead in quantum technologies and optimal-control protocols~\cite{rodriguez2024optimal,auffeves2022quantum,fellous2023optimizing}. We show that this large-charger requirement can be overcome by moving beyond purely population-driven charging to a coherence-assisted protocol.

While charger coherence and reservoir squeezing have previously been explored separately in QB schemes aimed at enhancing transient performance rather than ergotropy stabilization~\cite{centrone2023charging,li2025enhancing,lai2024quick}, this work presents the first study of their synergistic interplay in simultaneously catalyzing charging and enhancing robust steady-state ergotropy. We show that these two resources play distinct and complementary roles: initial charger coherence is the fundamental resource for attaining high-capacity, protected steady-state ergotropy, whereas reservoir squeezing provides a complementary coherence channel that dramatically accelerates the transient charging dynamics. Our framework further reveals that these advantages are mediated by the buildup of local battery coherence, which serves as the physical mechanism underlying both enhanced charging power and long-term work storage. 

Beyond these performance gains, our framework is characterized by its high experimental accessibility. In contrast to many coherence-assisted proposals requiring complex state engineering or large auxiliary systems~\cite{li2025enhancing,lai2024quick,mayo2022collective}, our approach utilizes spin-coherent states and squeezed vacuum reservoirs—resources routinely realized in modern quantum platforms. Specifically, spin-coherent states---typically prepared by applying a coherent driving field or resonant pulse to perform precise collective rotations---are accessible across diverse platforms, including Cavity QED~\cite{leroux2010implementation}, Bose–Einstein condensates~\cite{byrnes2015macroscopic}, NMR systems~\cite{auccaise2013spin,joshi2022experimental}, and nitrogen-vacancy centers~\cite{angerer2018superradiant}. These systems enable the operational integration of the coherent collective spins with a squeezed vacuum field, the generation of which is well-established in the microwave domain using Josephson or Traveling Wave Parametric Amplifiers~\cite{murch2013reduction, mallet2011quantum, toyli2016resonance}. This technical capability provides a direct pathway for implementing stabilized, coherence-driven protocols across the diverse frameworks currently at the forefront of experimental quantum battery research, including organic microcavities~\cite{quach2022superabsorption}, spin ensembles~\cite{niu2024experimental}, quantum dots~\cite{maillette2023experimental}, and superconducting circuits~\cite{hu2026quantum}. 




The remainder of this paper is organized as follows. In Sec.~\ref{sec:ergo_cohe}, we review the formal definitions of total, incoherent, and coherent ergotropy, establishing the $l_1$ norm of coherence as our primary metric for quantifying work-extraction capacity. Sec.~\ref{sec:system} introduces our charging scheme, specifying the charger’s initial state and the master equation governing the dissipative dynamics of the spin-ensemble system under a squeezed vacuum reservoir. In Sec.~\ref{sec:darkstate}, we analytically and numerically examine the dark-state manifold, demonstrating how initial charger coherence stabilizes the battery at a significantly higher steady-state ergotropy. Sec.~\ref{sec:dynamics} explores the time-resolved dynamics, identifying how both coherence resources accelerate the ergotropy charging speed, catalyze the superradiant scaling of the charging power, and, from a resource-accounting perspective, support a finite-time squeezing protocol. Finally, in Sec.~\ref{sec:conclusion}, we summarize our findings and discuss potential experimental realizations. 

\section{Ergotropy and Coherence}
\label{sec:ergo_cohe}
The work-extraction capability of the battery is defined as its ergotropy—the maximum work extractable through cyclic unitary operations~\cite{allahverdyan2004maximal}, considering a finite QB system coupled to a macroscopic work source. The battery dynamics is governed by a time-dependent Hamiltonian $\hat{H}(t) = \hat{H}_0 + \hat{V}(t)$, where $\hat{H}_0$ represents the free battery Hamiltonian and $\hat{V}(t)$ is the external driving or coupling to other systems, responsible for work exchange. To ensure cyclic evolution, we require $\hat{V}(0) = \hat{V}(\tau) = 0$, restoring the Hamiltonian to $\hat{H}_0$ at the end of the evolution at $\tau$. The resulting unitary evolution, $\hat{U}(t)=\mathcal{T}\exp\!\left[-{i}/{\hbar}\int_{0}^{t} \hat{H}(t')\,dt'\right]$ transforms the initial battery state $\rho_{B0}$ into $\rho_B(t) = \hat{U}(t) \rho_{B0} \hat{U}^\dagger(t)$ while preserving its spectrum. We consider the spectral decompositions of the initial configuration $\rho_{B0}$ and internal Hamiltonian $\hat{H}_0$: 
\begin{align}
\rho_{B0}=\sum_i r_i |\mathbf{r}_i\rangle\langle \mathbf{r}_i|,~~~H_0=\sum_j \epsilon_j |\mathbf{e}_j\rangle \langle \mathbf{e}_j |.
\label{rho0H0}
\end{align}
Here, $|\mathbf{r}_i\rangle$ and $|\mathbf{e}_j\rangle$ represent the eigenbasis of the initial battery state and the energy eigenbasis of the internal Hamiltonian, respectively. The corresponding eigenvalues are ordered such that $r_i \geq r_{i+1}$ and $\epsilon_j \leq \epsilon_{j+1}$. The battery ergotropy $W_B$ is then defined as:
\begin{align}
W_B = \mathrm{Tr}[\hat{H}_0 \rho_{B0}] - \min_{\hat{U}(\tau)}~\mathrm{Tr}[\hat{H}_0 \hat{U}(\tau) \rho_{B0} \hat{U}^\dagger(\tau)].
\label{ergomax}
\end{align}

The optimal transformation $\hat{U}_{m}(\tau)$ maps $\rho_{B0}$ to the passive state $\tilde{\rho}_{B}$, the minimum energy configuration reachable through unitary evolution. Since unitary operations preserve the spectrum of $\rho_{B0}$, the passive state must satisfy two requirements: (1) it is diagonal in the energy eigenbasis $\{|\mathbf{e}_j\rangle\}$, and (2) higher energy levels are monotonically less populated following the ordering established in Eq.~(\ref{rho0H0}). This configuration takes the form:
\begin{align}
\tilde{\rho}_B=\sum_k r_k |\mathbf{e}_k\rangle \langle \mathbf{e}_k |,
\label{passive}
\end{align} 
 When the battery state is mapped to this passive configuration, the ergotropy takes the standard form:
\begin{align}
W_B=\sum_{i,j}r_i \epsilon_j (|\langle \mathbf{r}_i|\mathbf{e}_j\rangle|^2-\delta_{ij}).
\label{ergomin}
\end{align}
Conversely, the violation of these two requirements for passivity reveals two distinct physical contributions to the ergotropy: (1) off-diagonal quantum coherence, and (2) population inversion among energy eigenstates. To quantify the work-extraction capacity residing in the off-diagonal elements, we adopt the $l_{1}$ norm of coherence~\cite{baumgratz2014quantifying}:

\begin{align}\label{l1co}
\mathcal{C}_{B} = \sum_{i \neq j} | \langle \mathbf{e}_i | \rho_{B0} | \mathbf{e}_j \rangle |.
\end{align}
To isolate the ergotropic contribution of coherence, we partition the total ergotropy into an incoherent (population inversion) part $W_B^P$ and a coherent part $W_B^C$~\cite{francica2020quantum}. This is achieved by defining the completely dephased state in the energy eigenbasis $\{|\mathbf{e}_j\rangle\}$ as $\rho_B^{\text{diag}} = \sum_j |\mathbf{e}_j\rangle\langle \mathbf{e}_j|\rho_{B0}|\mathbf{e}_j\rangle\langle \mathbf{e}_j|$. The two contributions to the ergotropy are then defined as:
\begin{align}
W_B^P:=W(\rho_B^{\mathrm{diag}}), \notag \\
W_B^C:=W_B-W_B^P.
\label{WPC}
\end{align}
This decomposition identifies $W_B^P$ as the work extractable from the diagonal population distribution alone and $W_B^C$ as the additional ergotropy enabled by off-diagonal coherence. 

\section{Dissipative Dynamics and Coherence Injection} \label{sec:system}
\begin{figure}[H]
  \centering
  \captionsetup{font=small}
  \hspace*{-0.1cm}\includegraphics[width=0.99\linewidth]{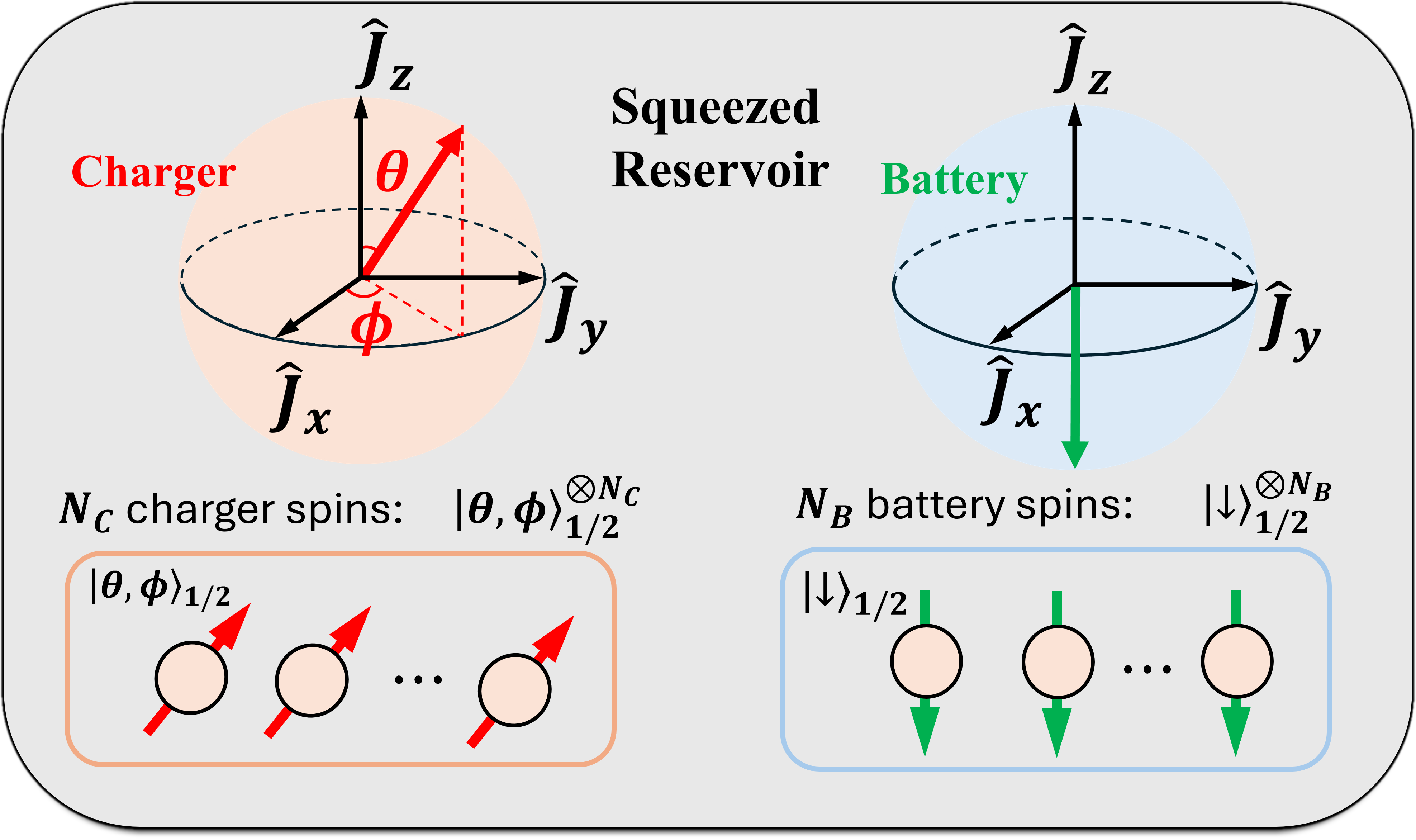}
  \caption{\justifying 
    Schematic of the model. A charger ($N_C$ spins) and a battery ($N_B$ spins) are coupled to a common reservoir. The charger is prepared in a product state $|\theta, \phi\rangle_{1/2}^{\otimes N_C}$, and the battery is initialized in the fully down-polarized state $|\downarrow \rangle^{\otimes N_B}$. The two large spheres represent the collective-spin states of the charger and battery, and the red and green arrows inside them indicate the corresponding collective spin orientations.
    }
  \label{fig:scheme}
\end{figure}

To realize the ergotropy stabilization and coherence-driven advantages, we introduce a physical framework for the dynamic injection of dual-channel coherence.  As shown in Fig.~\ref{fig:scheme}, our scheme models the charger and battery as two spin domains, $C$ and $B$, containing $N_C$ and $N_B$ identical spin-$1/2$ particles. These domains do not couple directly but interact solely through a shared dissipative reservoir, allowing the charger’s initial state and reservoir properties to be independently engineered. In standard open-system frameworks, environmental coupling typically constrains QB performance by dissipating the off-diagonal coherence and populations required for work extraction. To counteract these losses, we exploit \emph{dark states}, which are states immune to the dominant dissipation channels. This enables the battery to be simultaneously charged and stabilized against environmental decay~\cite{quach2020using}.


To mathematically describe the reservoir-mediated coupling, we characterize each domain $d \in \{C, B\}$ by collective spin operators with components $\hat J_{d}^\alpha = \sum_{j=1}^{N_d} \hat \sigma_{j,(d)}^{\alpha}/2$ for $\alpha \in \{x, y, z\}$, where $\hat \sigma_{j,(d)}^{\alpha}$ denotes the Pauli operator for the $j$th spin. The corresponding collective raising and lowering operators are $\hat J_d^{\pm} = \hat J_d^x \pm i\hat J_d^y$. Within the rotating-wave approximation, the interaction Hamiltonian between these collective spins and the bosonic reservoir modes is given by:
\begin{align}\label{Hamiltonian}
H_I(t) =& \hbar \sum_{\mathbf{k}}g_{\mathbf{k}} \left[  \hat a_{\mathbf{k}} (\hat J_C^+ + \hat J_B^+) e^{i(\nu_\mathbf{k} - \omega)t}\right.\cr
&\left. +  \hat a_{\mathbf{k}}^\dagger (\hat J_C^- + \hat J_B^-) e^{-i(\nu_\mathbf{k} - \omega)t} \right].
\end{align}
In Eq. \eqref{Hamiltonian}, $\hat a_{\mathbf{k}}$ and $\hat a_{\mathbf{k}}^{\dagger}$ denote the annihilation and creation operators for a reservoir mode with wave vector $\mathbf{k}$ and frequency $\nu_\mathbf{k}$, while $g_{\mathbf{k}}$ is the real-valued coupling strength. Physically, the dynamics are primarily governed by the resonant reservoir mode $\mathbf{k}_0$ satisfying the condition $\nu_{\mathbf{k}_0} = \omega$, where $\omega$ is the spin transition frequency. Within this resonance limit, as detailed in Appendix~\ref{appendixa} and following the standard treatment of squeezed reservoirs~\cite{agarwal1990cooperative}, the reduced dynamics of the spin system are described by the master equation:
\begin{align} \label{master3}
\dot{\rho} =& \gamma (\bar{n}+1)~(2\hat J^{-} \rho \hat J^+ -\{\hat J^+ \hat J^{-},~\rho\}) \notag \\
&+\gamma \bar{n}~(2\hat J^{+} \rho \hat J^{-} -\{\hat J^{-} \hat J^+,~\rho\}) \notag \\
&+\gamma\bar{m}^* \left( 2 \hat J^{-} \rho \hat J^{-} -\{\hat J^{-} \hat J^{-},~\rho\} \right) \notag\\
&+\gamma\bar{m} \left( 2 \hat J^+ \rho  \hat J^+ -\{\hat J^+ \hat J^+,~\rho\}  \right)  .
\end{align}
In this expression, $\hat{J}^\pm \equiv \hat{J}_C^\pm + \hat{J}_B^\pm$ represent the collective system operators, and $\gamma$ characterizes the effective decay rate. The terms $\bar{n} = \langle \hat{a}_{\mathbf{k}_0}^\dagger \hat{a}_{\mathbf{k}_0} \rangle$ and $\bar{m} = \langle \hat{a}_{\mathbf{k}_0} \hat{a}_{\mathbf{k}_0} \rangle$ denote the mean photon number and the anomalous correlation of the resonant reservoir mode, respectively.

With the system dynamics established, we now specify the conditions required to drive the coherence-injection process. As discussed in Sec.~\ref{sec:ergo_cohe}, QB ergotropy depends crucially on its quantum coherence. In our model (Fig.~\ref{fig:scheme}), we move beyond the standard fully excited charger by initializing the charger in a coherent superposition, allowing for coherence-driven charging throughout the dissipative evolution. At the microscopic level, each charger spin is prepared in a coherent state $|\theta,\phi\rangle_{1/2}=e^{-i\phi/2}\cos(\theta/2) ~\ket{\uparrow} + e^{i\phi/2}\sin(\theta/2) ~\ket{\downarrow}$, where $\theta$ and $\phi$ denote the polar and azimuthal angles on the Bloch sphere, respectively. For the collective ensemble of $N_C$ identical spins $\ket{\theta,\phi}_{1/2}^{\otimes N_C}$, this preparation corresponds to a collective spin-coherent state within the symmetric subspace ($J=N_C/2$)~\cite{radcliffe1971some,arecchi1972atomic}:
\begin{align}\label{spco}
  \ket{\theta,\phi}_C
  = e^{-i\phi \hat J_C^z}\,e^{-i\theta \hat J_C^y}\,\ket{J,J}_C,
\end{align}
where $\ket{J,J}_C$ denotes the fully excited Dicke state of the charger. Conversely, the battery is initialized in the fully down-polarized state $|\downarrow\rangle^{\otimes N_B}$ at $t=0$, an assumption maintained throughout this work. 

The spin-coherent state $\ket{\theta,\phi}_C$ is a superposition in the Dicke basis with coherence among the Dicke levels, the details are shown in Appendx~\ref{appendixb}. the angle $\theta$ plays a dual role: while increasing $\theta$ generates the internal coherence required for ergotropy stabilization, it simultaneously reduces the initial excitation energy of the charger. This establishes a fundamental trade-off between coherence-assisted and population-driven charging, which we analyze in the following sections.


In addition to the coherence injected by the charger’s initial state, the reservoir itself serves as an external source of coherence via quantum squeezing~\cite{centrone2023charging}. A squeezed vacuum introduces correlated two-photon processes governed by the squeezing operator $S = \prod_{\mathbf{k}} \exp[ (\xi_\mathbf{k} {{\hat a}_\mathbf{k}^{\dagger2}} - \xi_\mathbf{k}^* \hat a_\mathbf{k}^2 )/2]$, where the parameter $\xi_\mathbf{k} = r(\mathbf{k})e^{i\varphi(\mathbf{k})}$ characterizes the squeezing amplitude and phase. Focusing on the resonant mode $\mathbf{k}_0$ and using the definitions $r \equiv r(\mathbf{k}_0)$ and $\varphi \equiv \varphi(\mathbf{k}_0)$, the parameters in Eq.~(\ref{master3}) satisfy $\bar{n} = \sinh^2(r)$ and $\bar{m} = \sinh(r) \cosh(r) e^{i\varphi}$. In the pure squeezed vacuum limit, where the identity $\bar{n}(\bar{n}+1) = |\bar{m}|^2$ holds, Eq.~\eqref{master3} can be recast into the compact form: 
\begin{equation}
\dot{\rho} = \gamma\left(2 \hat L_{\varphi} \rho \hat L_{\varphi}^\dagger - \hat L_{\varphi}^\dagger \hat L_{\varphi} \rho - \rho \hat L_{\varphi}^\dagger \hat L_{\varphi}\right),
\label{dynamical2}
\end{equation}
the associated jump operator $\hat L_{\varphi}$ is defined as:
\begin{equation}
\hat L_{\varphi} = \hat J^{-} \cosh(r) e^{ \frac{i\varphi}{2}} + \hat J^+ \sinh(r)e^\frac{-i \varphi}{2}.
\label{eq:jump_op}
\end{equation}
We note that $\hat{L}_{\varphi}$ also depends on the squeezing parameter $r$, though we leave this dependence implicit for notational brevity. One can readily verify that this operator satisfies the commutation relation $[\hat L_{\varphi}, \hat L_{\varphi}^\dagger] = [\hat J^{-}, \hat J^+]$. 

The performance of the charger-battery system is governed by several primary control parameters: the charger polar angle $\theta$ and azimuthal angle $\phi$, alongside the reservoir squeezing strength $r$ and phase $\varphi$. While the system dynamics depend non-trivially on $\theta$, the role of the two optical phases, $\phi$ and $\varphi$, are dictated by symmetry. Specifically, for either an unsqueezed reservoir ($r=0$) or a coherence-free charger ($\theta=0$), the system remains rotationally invariant, rendering the corresponding phase parameters physically irrelevant. This symmetry is broken only when both finite charger coherence $(\theta \neq 0)$ and reservoir squeezing $(r \neq 0)$ coexist. In this regime, $\phi$ and $\varphi$ define a shared transverse reference frame. As derived in Appendix~\ref{appendixb}, the dynamics then depend solely on the relative phase:   
\begin{align}
\delta \equiv \varphi - 2\phi.
\end{align}
This identifies $\delta$ as the tuning parameter representing the mutual alignment between the charger initial coherence and the reservoir squeezing axis; consequently, the resulting ergotropy and energy are functions of $\delta$, $\theta$ and $r$. 

\section{Long-Term Stabilization under Dark-State Protection}\label{sec:darkstate}
Building on the established charging scheme, we now characterize the resulting steady-state ergotropy and the structure of the protected dark-state manifold. While joint coupling of the charger-battery system to a common reservoir is known to stabilize the energy of open QBs~\cite{quach2020using}, we demonstrate how the integration of internal charger coherence and external reservoir squeezing fundamentally modifies these steady states to maximize extractable work. We begin in Subsec.~\ref{subsec:twospin} by analytically examining a minimal two-spin system ($N_C = N_B = 1$) to identify the explicit form of the steady-state solutions and their associated ergotropy. In Subsec.~\ref{subsec:multispin}, we extend the analysis to multi-spin systems, where both analytical and numerical results reveal that coherence modifies the dark states and enhances the QB ergotropy. In Subsec.~\ref{subsec:entang}, we show that entanglement between the charger and the battery suppresses the coherence-related part of the QB ergotropy.

\subsection{Minimal Two-Spin System}
\label{subsec:twospin}
We first analyze the minimal system consisting of one charger spin and one battery spin ($N_C=N_B=1$). The steady states for this configuration are determined by the dark-state condition $\hat{L}_{\varphi}|\Psi_{d_{i}}\rangle=0$, yielding two linearly independent solutions $\{|\Psi_{d_{i}}\rangle\}$ that remain immune to environmental dissipation (see Appendix~\ref{appendixc} for explicit derivations). While these dark states define the possible stationary configurations, the final state reached by the system depends critically on the charger’s initial preparation. Specifically, we consider a single-spin charger prepared in the spin-coherent state:
\begin{align}
\ket{\theta,\phi}_C =e^{-\frac{i\phi}{2}}\cos\left(\frac{\theta}{2}\right) ~\ket{\uparrow}_C + e^{\frac{i\phi}{2}} \sin\left(\frac{\theta}{2}\right) ~\ket{\downarrow}_C.
\end{align}
The analytical steady state of the charger-battery system, evolving from the initial state $|\theta,\phi\rangle_C|\downarrow\rangle_B$, is derived in Appendix~\ref{appendixc}. By tracing out the charger degrees of freedom, the reduced density matrix of the battery spin $\rho_B$ in the energy eigenbasis $\{|\downarrow\rangle_B, |\uparrow\rangle_B\}$ is given by 
\begin{align} \label{densi_co}
\rho_B =& \frac{1}{8(1+2\bar{n})} \nonumber \\
&\scalebox{0.8}{$\times\begin{pmatrix}
7 + 8\bar{n} - \cos\theta & -2\sin\theta (1+\bar{n}  + |\bar{m}|e^{-i\delta}) \\
-2\sin\theta (1+\bar{n} + |\bar{m}|e^{i\delta}) & 1 + 8\bar{n} + \cos\theta
\end{pmatrix}$},
\end{align}
where the parameters $\bar{n}$ and $\bar{m}$ are defined below Eq.~(\ref{master3}). The battery energy $E_B$ is 
\begin{eqnarray}\label{energy_co}
\frac{E_B}{\hbar\omega}=-\frac{3 - \cos\theta}{8(1+2\bar n)} +\frac{1}{2}.
\end{eqnarray}
Equations (\ref{densi_co}) and (\ref{energy_co}) represent the general steady-state solution for the minimal two-spin system, where battery performance is jointly governed by internal charger coherence ($\theta$) and external reservoir squeezing ($r$). The resulting ergotropy is given by Eq.~(\ref{ergo_co}), the non-negativity of which is rigorously established in Appendix~\ref{appendixc}. This unified expression shows that the work-extraction capacity is dictated by the mutual alignment between the charger's internal state and the reservoir's external coherence channel. Notably, for this minimal system, the population distribution in Eq. (\ref{densi_co}) never reaches inversion regardless of the parameters. Consequently, the incoherent ergotropy component is identically zero ($\mathcal{W}_B^P = 0$), and all extractable work originates solely from the local battery coherence. This confirms that in the single-spin limit, internal charger coherence is not merely an enhancement but the indispensable resource for ergotropy storage. 

To isolate the distinct physical mechanisms driving this performance, we now examine two fundamental limiting regimes. When the charger is prepared without initial coherence ($\theta=0$) but with squeezing ($\bar{n}\neq 0$), the battery steady state $\rho_B$ [Eq.~(\ref{densi_co})] remains diagonal in the energy eigenbasis. In this regime, although the squeezed vacuum reservoir introduces external coherence that boosts the stored energy, it cannot independently generate the local battery coherence necessary for ergotropy extraction. Consequently, the extractable work vanishes ($W_B=0$), as can be verified by taking the $\theta=0$ limit of Eq. (\ref{ergo_co}). Conversely, in the vacuum reservoir limit ($\bar{n} = \bar{m} = 0$), the ergotropy simplifies to 
\begin{align}
W_B = \frac{-3 + \cos\theta + \sqrt{10 - 6\cos\theta + 3\sin^2\theta}}{8},
\end{align}
which vanishes at $\theta=0$ and is strictly positive for any $\theta > 0$. In the $\theta > 0$ regime, the battery energy decreases relative to the $\theta=0$ case because the charger initially contains less excitation. However, its preparation in a spin coherent state induces local battery coherence, generating the elements in the off-diagonal necessary for finite ergotropy. This finite ergotropy in the vacuum reservoir case originates entirely from the coherence present in the initial state of the charger.

Beyond these limits, the full analytical expression for the ergotropy [Eq.~(\ref{ergo_co})] reveals the general dependencies that dictate performance in the $N_C=N_B=1$ system. First, the extractable work is maximized under the phase-matching condition $\delta = 0$, where the internal charger coherence is optimally aligned with the reservoir's squeezing axis. Second, the ergotropy exhibits a non-monotonic dependence on the polar angle $\theta$: while increasing $\theta$ injects the coherence necessary for work extraction, it simultaneously reduces the initial excitation energy. For example, at $r=0.5$ and $\delta=0$, one finds an optimal angle $\theta \approx 0.46\pi$, at which the battery ergotropy is maximized. As we examine in the following section, these single-spin trends---the optimality of $\delta=0$ and the resource trade-off in $\theta$---remain fundamentally robust and serve as the prototypical basis for the collective advantages observed in larger multi-spin ensembles.

\subsection{Multi-spin system}
\label{subsec:multispin}
While the results for the minimal two spin case provide physical intuition, the operational advantages of collective charging such as superradiant scaling emerge most clearly in larger ensembles. We therefore extend our analysis to multi-spin systems. Analytical expressions are derived for small-scale systems up to the case of $N_C = N_B = 2$ (Appendix~\ref{appendixc}), while the dynamics of larger spin-number configurations are solved numerically (Appendix~\ref{appendixd}). To quantify this collective performance, we introduce the normalized energy per battery spin, $\mathcal{E}_{B}=E_{B}/(N_{B}\hbar\omega)$, and the normalized ergotropy per battery spin, $\mathcal{W}_{B}=W_{B}/(N_{B}\hbar\omega)$. 

\begin{figure}[H]
  \centering
  \captionsetup{font=small}
  \hspace*{-0.3cm}\includegraphics[width=1.03\linewidth]{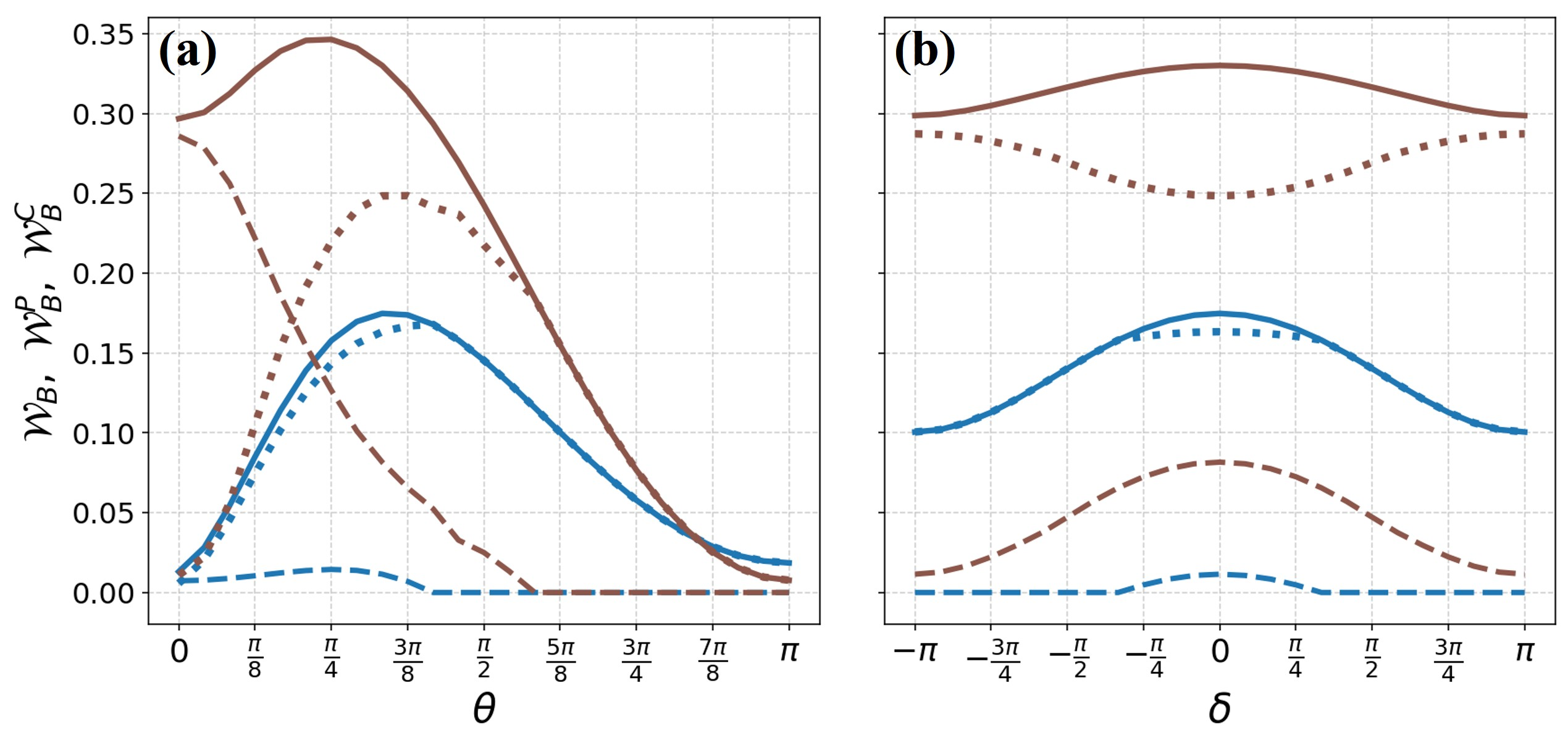}
  \caption{\justifying Steady-state battery performance in multi-spin systems for $N_B = 4$ and various charger sizes: $N_C = 4$ (blue) and $N_C = 8$ (brown). The squeezed vacuum parameter is fixed at $r = 0.5$. Solid, dotted, and dashed curves represent total ergotropy $\mathcal{W}_B$, coherent ergotropy $\mathcal{W}_B^C$, and population-driven ergotropy $\mathcal{W}_B^P$, respectively. (a): Dependence on the charger polar angle $\theta \in [0, \pi]$ with the relative phase fixed at $\delta = 0$. (b): Dependence on the relative phase $\delta \in [-\pi, \pi]$ with the charger polar angle fixed at $\theta = \pi/3$.
  }
  \label{ssfigphase}
\end{figure}

To investigate the physical conditions required to optimize battery performance, we first lean on the analytical foundations of the single-spin case. Specifically, Eq.~(\ref{ergo_co}) establishes $\delta = 0$ as the optimal phase-matching condition for $N_C = N_B = 1$. We therefore fix the relative phase at this optimal value to examine the $\theta$-dependence across various system sizes in Fig.~\ref{ssfigphase}(a). As illustrated, increasing $\theta$ from small to moderate values reveals a fundamental resource trade-off: it enhances the coherent ergotropy $\mathcal{W}_B^C$ by injecting charger coherence but simultaneously depletes the population-driven component $\mathcal{W}_B^P$ by reducing initial excitation energy. When $\theta$ exceeds a specific threshold, $\mathcal{W}_B^P$ vanishes as the charger lacks the energy to maintain population inversion in the battery. In contrast, $\mathcal{W}_B^C$ remains finite throughout the entire range $[0, \pi]$ of $\theta$. 
Crucially, the multi-spin architecture enables finite ergotropy at both the incoherent ($\theta=0$) and ground-state ($\theta=\pi$) limits—features fundamentally forbidden in $N_C = N_B = 1$ case. As proved analytically in Appendix~\ref{appendixc}, these collective advantages stem from the initial state's overlap with dark-state cross-terms, a mechanism unique to $N_B > 1$ architectures. While the ergotropy at $\theta=\pi$ arises solely from reservoir-induced correlations, the non-vanishing ergotropy at $\theta=0$ highlights a purely architectural pathway for extraction that remains robust even in the absence of initial charger coherence. For $N_C = N_B = 4$, the total ergotropy $\mathcal{W}_B$ reaches its maximum at $\theta \approx \pi/3$, representing an optimal balance between coherent and population-driven charging.




Having identified the optimal polar angle, we now verify if the $\delta = 0$ condition remains robust for multi-spin ensembles in Fig.~\ref{ssfigphase}(b). As illustrated, the incoherent contribution $\mathcal{W}_{B}^{P}$ is consistently maximized at $\delta=0$ regardless of system size. In contrast, the coherent contribution $\mathcal{W}_{B}^{C}$ attains a maximum at $\delta=0$ for smaller spin-number ratios $N_{C}/N_{B}$, but a minimum for larger ratios. Nevertheless, the total ergotropy $\mathcal{W}_{B}$ remains strictly optimal at $\delta=0$ across all ratios. 
We therefore fix both $\theta = \pi/3$ and $\delta = 0$ for the subsequent multi-spin analysis.

\begin{figure}[H]
  \centering
  \captionsetup{font=small}
  \hspace*{-0.3cm}\includegraphics[width=1.03\linewidth]{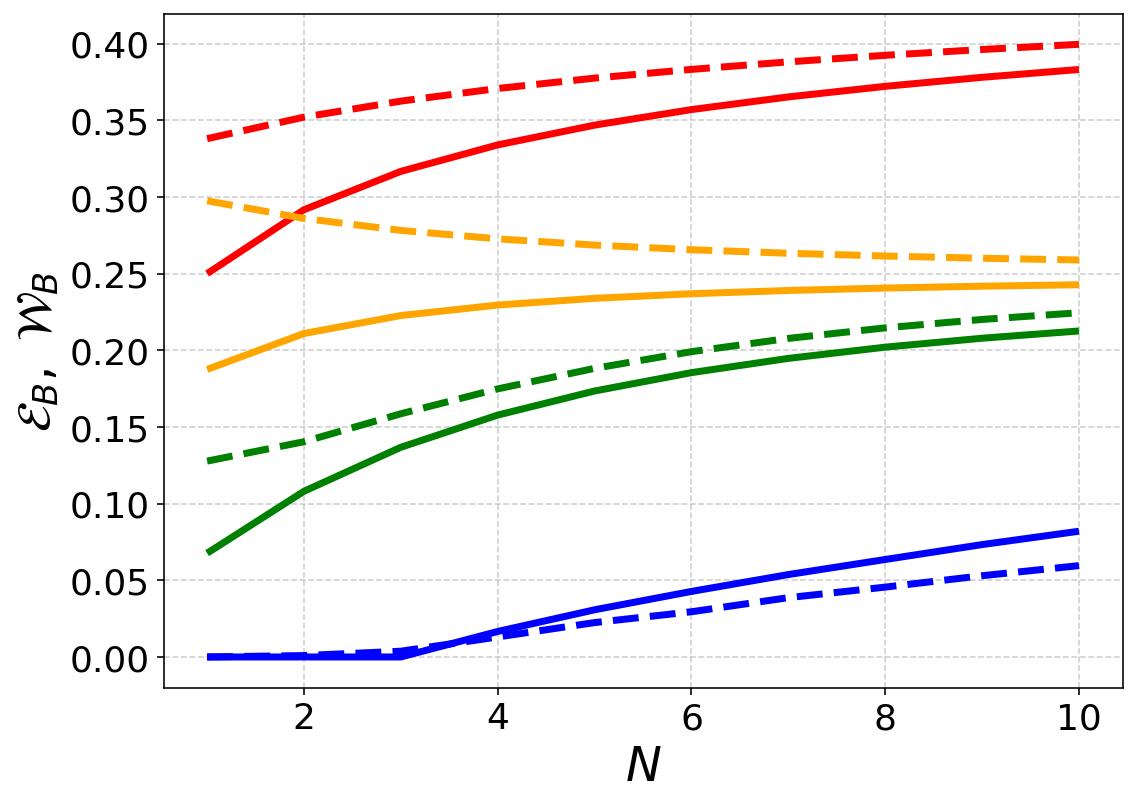}
  \caption{\justifying Scaling of steady-state battery performance with system size $N$, under a fixed spin ratio $N_C = N_B = N$. Normalized battery energy $\mathcal{E}_B$ (red and yellow) and ergotropy $\mathcal{W}_B$ (blue and green) as functions of $N_B$ for vacuum (solid) and squeezed-vacuum ($r = 0.5,$ dashed) reservoirs. Specifically, red ($\mathcal{E}_B$) and blue ($\mathcal{W}_B$) curves correspond to the all-spin-up charger ($\theta = 0$), while yellow ($\mathcal{E}_B$) and green ($\mathcal{W}_B$) curves denote the spin-coherent charger ($\theta = \pi/3$). All results are evaluated at the relative phase $\delta = 0$.
  }
  \label{ssfigN}
\end{figure}

To further explore how initial-state coherence and reservoir squeezing jointly influence the steady-state performance of quantum batteries, we numerically compute the steady-state solutions as a function of the system size $N$, maintaining a fixed spin ratio $N_C = N_B = N$. Figure~\ref{ssfigN} displays the normalized battery energy $\mathcal{E}_B$ and ergotropy $\mathcal{W}_B$ for both vacuum and squeezed-vacuum reservoirs. The charger's polar angle is held at either the incoherent all-spin-up state ($\theta = 0$) or a coherent state $(\theta = \pi/3)$. For either charger type, $\mathcal{E}_B$ remains comparable for both reservoirs, though it is lower for the spin-coherent charger due to its reduced initial excitation.

The ergotropy per spin $\mathcal{W}_B$ is consistently smaller than the total energy $\mathcal{E}_B$ for any given configuration. For the coherence-free charger ($\theta = 0$), $\mathcal{W}_B$ increases with $N_B$ after a distinct onset across all reservoirs. Under vacuum and squeezed-vacuum reservoirs, initializing the charger in a spin-coherent state substantially enhances the steady-state ergotropy compared to the $\theta = 0$ case, following a similar scaling trend. Notably, the benefit of squeezing over vacuum is size-dependent and diminishes as the system scales. Overall, while the coherent charger carries lower internal energy $\mathcal{E}_B$ due to reduced initial excitation, it yields significantly higher extractable work $\mathcal{W}_B$ in vacuum and squeezed environments compared to the $\theta = 0$ case. This confirms that initial charger coherence, rather than reservoir squeezing, is the critical resource for maximizing steady-state QB ergotropy.

\subsection{The Role of Entanglement} \label{subsec:entang}
To further understand the relationship between quantum correlations and work extraction, we analyze the entanglement between the charger and battery domains. While entanglement serves as a beneficial \cite{alicki2013entanglement, binder2015quantacell} or neutral \cite{ghosh2021fast, hovhannisyan2013entanglement} thermodynamic resource in several unitary charging protocols, its role in shared-dissipative frameworks is qualitatively distinct. We employ the logarithmic negativity to quantify this entanglement \cite{vidal2002computable, plenio2005logarithmic}, defined as
\begin{align}\label{negat}
\mathcal{S}_B=\log_2||\rho^{\Gamma_B}||,
\end{align}
where $\Gamma_B$ denotes the partial transpose with respect to the battery subsystem $B$.

\begin{figure}[H]
  \centering
  
  \captionsetup{font=small}
  \hspace*{-0.3cm}\includegraphics[width=1.03\linewidth]{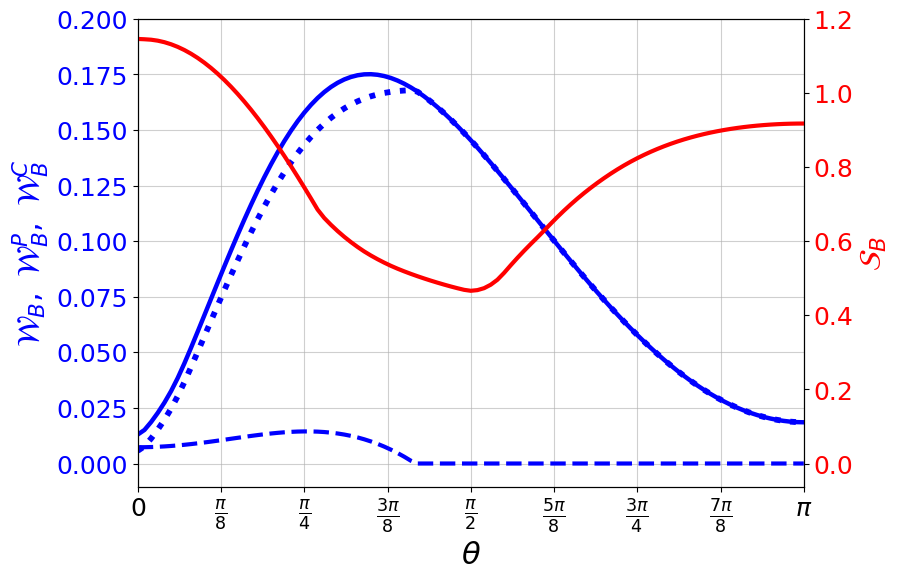}
  \caption{\justifying Steady-state ergotropy components and subsystem entanglement as functions of the charger polar angle $\theta$. Total ergotropy $\mathcal{W}_B$ (blue solid), population contribution $\mathcal{W}_B^P$ (blue dashed) and coherence-induced contribution $\mathcal{W}_B^C$ (blue dotted) are contrasted against the battery-charger entanglement $\mathcal{S}_B$ (red solid). Results are calculated for a squeezed-vacuum ($r = 0.5$) reservoir with spin numbers $N_B = N_C = 4$ at the relative phase $\delta=0$. } 
  \label{fig:loganeg}
\end{figure}

The ergotropy components ($\mathcal{W}_B$, $\mathcal{W}_B^P$, $\mathcal{W}_B^C$) and the logarithmic negativity ($\mathcal{S}_B$) as functions of the charger’s polar angle $\theta$ are shown in Fig.~\ref{fig:loganeg}. The non-smooth ‘kinks’ observed in these metrics are physical features resulting from the spectral reordering required to define the passive state. Specifically, as $\theta$ varies, crossings in the energy level populations induce discrete changes in the sorting permutation to maintain the monotonic requirement. This results in a discontinuous first derivative of the passive state energy, which directly manifests as the observed non-smooth features in the ergotropy metrics. Similarly, the kinks in $\mathcal{S}_B$ occur when an eigenvalue of the partial transpose crosses zero.

In the present shared-dissipative setting, Fig.~\ref{fig:loganeg} shows that the battery-charger entanglement $S_B$ generally competes with the battery's coherence-based work-storage capability. Stronger charger-battery entanglement makes the battery's reduced state more mixed, which in turn suppresses the local coherence and purity available for work extraction. As a result, part of the system's useful quantum character is redistributed into nonlocal correlations rather than retained locally in the battery. The relation between $S_B$ and the coherence-induced ergotropy $W_B^C$, however, is not strictly one-to-one, because $W_B^C$ depends not only on local coherence but also on the spectral structure of the reduced battery state. In particular, while the local battery coherence is maximized near $\theta=\pi/2$, $W_B^C$ reaches its maximum at a slightly smaller angle, because increasing $\theta$ injects more coherence but also narrows the diagonal population profile of the reduced battery state, thereby limiting the ergotropy upper bound accessible from coherence alone. Overall, these results indicate that, in our scheme, the quantum advantage is tied more closely to coherent cooperative interactions than to the buildup of bipartite entanglement.

Our steady-state analysis has shown that initial charger coherence is the dominant factor for final ergotropy, whereas reservoir squeezing offers diminishing utility in the steady-state limit. Having identified these stationary performance limits, Sec.~\ref{sec:dynamics} next investigates the time-resolved evolution of these resources to determine how reservoir squeezing and initial charger coherence collaboratively provide a transient advantage and enhance the overall charging performance.

\section{Accelerated Charging and Resource Optimization} \label{sec:dynamics}
\subsection{Time-Resolved Evolution and Power Scaling}

While the steady-state analysis identifies the ultimate energy, coherence, and ergotropy achievable under different reservoir conditions and initial states, it does not reveal how quickly these quantities build up or decay during the charging process. To address this, we numerically solve the master equation~(\ref{master3}) and examine the time-resolved evolution of the battery’s ergotropy and coherence, using the numerical framework described in Appendix~\ref{appendixd}. This analysis shows that reservoir squeezing and initial charger coherence each independently accelerate the early-time buildup of work-extraction capacity, while their combination produces the strongest enhancement and most effectively mitigates the dephasing that would otherwise reduce the stored ergotropy before long-term dark-state stabilization is reached.

\begin{figure}[ht]
  \centering
  \captionsetup{font=small}
  \hspace*{-0.3cm}\includegraphics[width=1.0\linewidth]{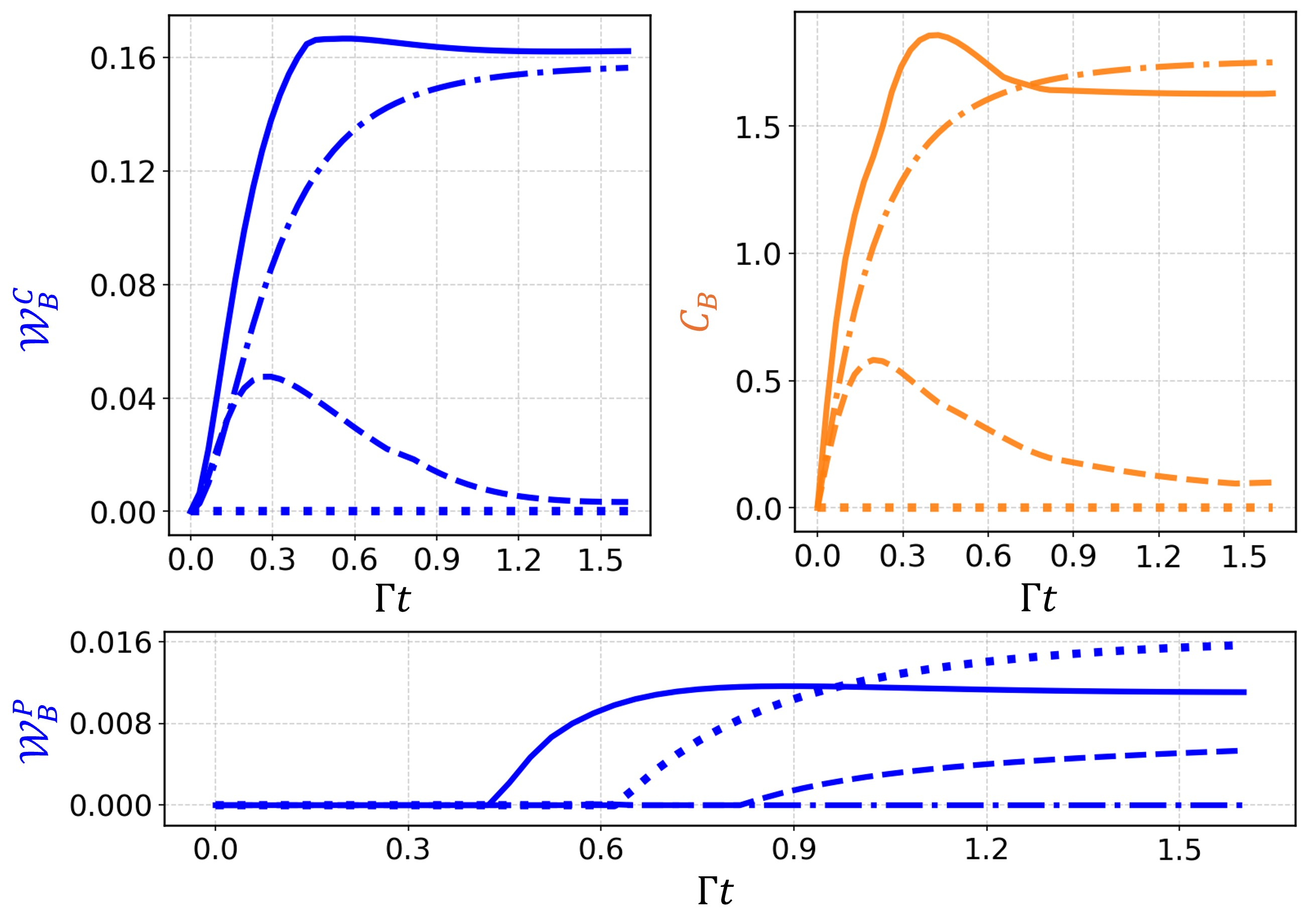}  
  \caption{\justifying Time evolution of the coherent ergotropy $\mathcal{W}_{B}^{C}$ (top panel, blue), incoherent ergotropy $\mathcal{W}_{B}^{P}$ (bottom panel, blue), and the local coherence $\mathcal{C}_{B}$ (top panel, orange) under different reservoir and initial state conditions: vacuum reservoir with all-spin-up charger ($\theta=0$) (dotted), squeezed-vacuum ($r=0.5$) reservoir with all-spin-up charger ($\theta=0$) (dashed), vacuum reservoir with spin-coherent charger ($\theta=\pi/3$) (dot-dashed), and squeezed-vacuum ($r=0.5$) reservoir with spin-coherent charger ($\theta=\pi/3$) (solid). Other parameters used are $N_{C}=N_{B}=4$ and $\delta=0$.}
  \label{dy_P:a}
\end{figure}
Figure~\ref{dy_P:a} illustrates the time evolution of the battery ergotropy components $\mathcal{W}_B^C$ and $\mathcal{W}_B^P$ alongside the local coherence $\mathcal{C}_B$ for $N_C = N_B = 4$. 
Notably, when charger and battery sizes are comparable, achieving significant population inversion is fundamentally constrained; thus, charging performance in this regime is governed by the coherent ergotropy $\mathcal{W}_B^C$. As shown in the figure, $\mathcal{W}_B^C$ and $\mathcal{C}_B$ exhibit synchronous evolution across all studied configurations, confirming that the buildup of coherent ergotropy is fundamentally driven by the generation of local battery coherence. For an all-spin-up charger ($\theta=0$) in a vacuum reservoir, both $\mathcal{C}_B$ and $\mathcal{W}_B^C$ remain strictly zero throughout the interaction. Conversely, introducing either the external coherence channel via reservoir squeezing or the internal coherence channel via a spin-coherent charger ($\theta =\pi/3$) triggers battery coherence immediately upon the onset of the interaction. This gives rise to an instantaneous work-extraction capability and an immediate increase in $\mathcal{W}_B^C$, in contrast to the delayed onset of $\mathcal{W}_B^P$.

\begin{figure}[H]
  \centering  
  \captionsetup{font=small}
  \hspace*{-0.3cm}\includegraphics[width=0.95\linewidth]{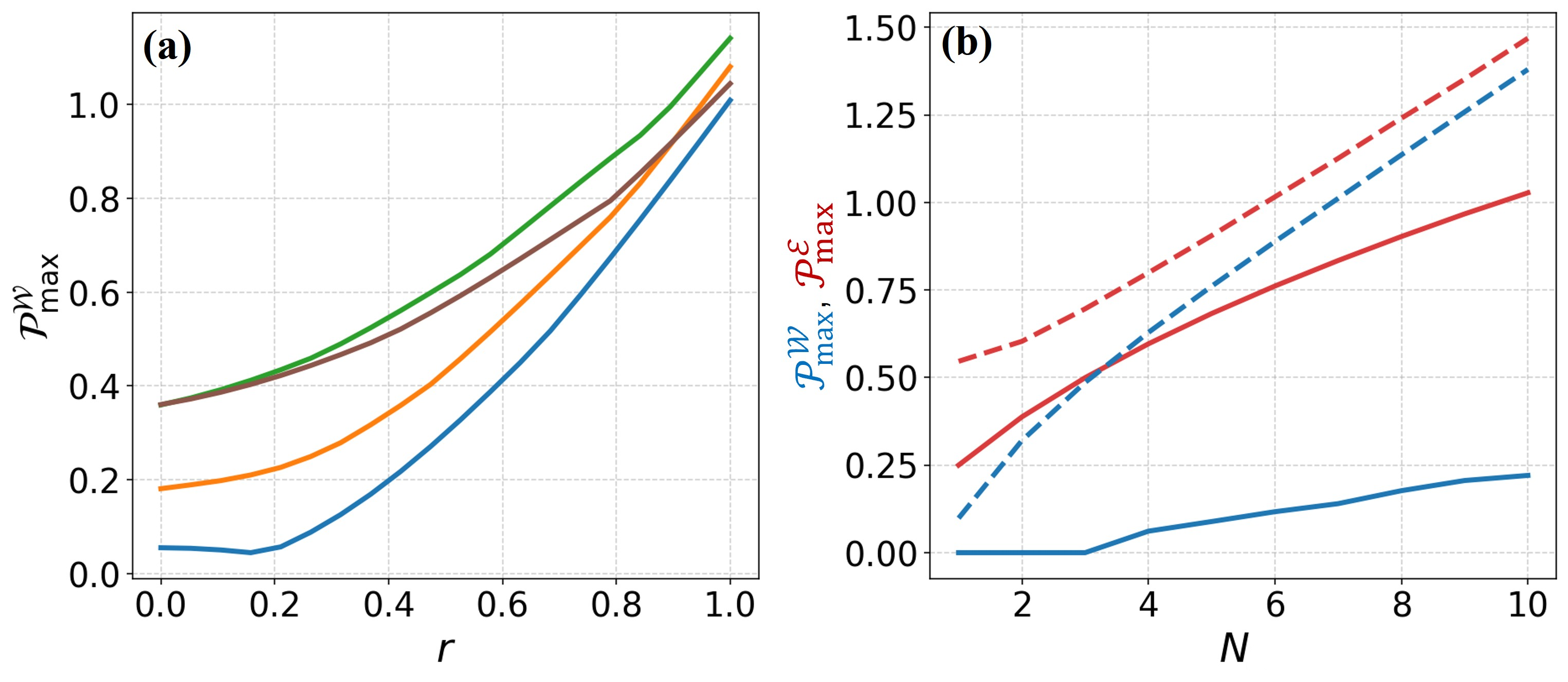}
  \caption{\justifying Scaling of maximum charging power per spin. (a): Maximum ergotropy charging power $\mathcal{P}_{\max}^{\mathcal{W}}$ as a function of the squeezing parameter $r$ for various charger polar angles $\theta$. Blue, orange, green, and brown curves denote $\theta = 0$, $\pi/6$, $\pi/3$, and $\pi/2$, respectively. Other parameters used are $N_C=N_B=4$ and $\delta=0$. (b): Maximum energy charging power $\mathcal{P}_{\max}^{\mathcal{E}}$ (red) and maximum ergotropy charging power $\mathcal{P}_{\max}^{\mathcal{W}}$ (blue) as functions of system size $N$ under $N_C=N_B=N$. Solid curves represent the baseline vacuum reservoir ($r=0$) with an all-spin-up charger ($\theta=0$); dashed curves represent the squeezed-vacuum ($r=0.5$) reservoir with a spin-coherent charger ($\theta=\pi/3$). Other parameter used is $\delta = 0$.  } 
  \label{dy_P}
\end{figure}
Beyond these correspondences, Fig.~\ref{dy_P:a} highlights how internal and external coherence jointly shape the overall charging trajectory. For any fixed reservoir configuration, the spin-coherent charger configuration yields significantly higher peak and steady-state ergotropy than the all-spin-up alternative. This enhancement is reflected in the initial slopes of the curves in Fig.~\ref{dy_P:a}, which directly correspond to the maximum charging powers analyzed later in Fig.~\ref{dy_P}. For a fixed initial charger state, a comparison between the squeezed-vacuum and vacuum reservoirs shows that squeezing induces a faster buildup of both $\mathcal{W}_B^C$ and $\mathcal{C}_B$ immediately after the interaction begins. However, after reaching a maximum, $\mathcal{W}_B^C$ and $\mathcal{C}_B$ in the squeezed-vacuum case gradually decay toward their steady-state values for both charger types. This occurs because dissipation channels inevitably suppress coherence over time. The transition from the peak value to the final steady state can therefore be interpreted as an effective discharging process, in which useful ergotropy is lost through dephasing before the system becomes protected by dark states. For the all-spin-up charger, this process results in the eventual loss of all $\mathcal{W}_B^C$. Conversely, when the charger is initially spin-coherent, this discharging process is strongly suppressed by the dark-state protection, allowing the battery to stabilize at a finite value of $\mathcal{W}_B^C$. Consequently, combining reservoir squeezing with charger-state coherence yields the optimal charging performance.

To further investigate the effect of squeezing and charger initial coherence on charging dynamics, we define the total ergotropy and energy charging power per battery spin as
\begin{align}\label{Pdef}
\mathcal{P}^\mathcal{W}=\frac{d\mathcal{W}_B}{dt}, \quad \mathcal{P}^\mathcal{E}=\frac{d\mathcal{E}_B}{dt}.
\end{align}     
We first analyze the maximum ergotropy charging power, $\mathcal{P}_{\max}^{\mathcal{W}} = \mathrm{Max}_t \mathcal{P}^{\mathcal{W}}(t)$, as a function of the squeezing parameter $r$ for various charger polar angles $\theta$. As shown in Fig.~\ref{dy_P}(a), the charging power grows monotonically with increasing $r$, confirming that reservoir squeezing accelerates the work-extraction rate. However, its dependence on the charger preparation is nonmonotonic: increasing $\theta$ initially enhances $\mathcal{P}_{\max}^{\mathcal{W}}$ by injecting coherence, but beyond an optimal range, the reduction in initial excitation energy begins to dominate, causing the charging power to decrease. 

We next evaluate the collective scaling behavior by analyzing the maximum charging power as a function of the system size $N$. As illustrated in Fig.~\ref{dy_P}(b), under the symmetric spin-ratio constraint $N_C = N_B = N$, the charging power per spin increases linearly with $N$, implying that the total charging power follows a superradiant $N^2$ enhancement. We emphasize that this scaling is obtained at fixed charger-to-battery ratio, so that the collective advantage is assessed within a resource-consistent comparable-size regime rather than by enlarging only one subsystem. Notably, this quadratic scaling is universal, persisting regardless of whether coherence or squeezing is introduced. This advantage is manifest in both energy and ergotropy charging regimes.  Furthermore, the simultaneous application of reservoir squeezing and a spin-coherent charger not only boosts the absolute magnitude of charging power but also steepens the slope of the $\mathcal{P}$--$N$ relationship ($d\mathcal{P}^{\mathcal{E},\mathcal{W}}_{\max}/dN$).  This demonstrates that these quantum-coherent resources do not merely enhance individual charging events but actively catalyze the collective advantage, enabling the system to harness superradiant emission more effectively from the very onset of the interaction.

Our dynamical analysis shows that reservoir squeezing and initial charger coherence accelerate the charging process most effectively when used together. In the comparable charger-battery size regime, this enhancement is driven by the rapid buildup of local battery coherence $\mathcal{C}_B$ and the corresponding coherent ergotropy $\mathcal{W}_B^C$, which bypass the delayed onset of population-driven charging. Importantly, this coherence-assisted advantage remains fully compatible with the universal $N^2$ superradiant scaling of charging power.

\subsection{Resource Optimization with Finite-Time Squeezing Protocol}
The dynamical advantages identified above also motivate a closer examination of resource cost. In particular, one must account for the energetic overhead associated with preparing the coherent charger and sustaining the squeezed reservoir. Compared to earlier stabilization mechanism~\cite{quach2020using} which utilize population-driven charging from fully excited sources, initializing the charger in a spin-coherent state $|\theta, \phi\rangle$ follows an operational protocol similar to a standard all-spin-up preparation. In most experimental platforms, both are typically realized via a single resonant pulse, implying that the stabilization benefits of charger coherence can be achieved using standard control sequences without necessitating the complex state-engineering required by other coherence-assisted protocols.

The primary challenge regarding resource accounting in reservoir-engineered charging protocols is that the continuous maintenance of a squeezed vacuum constitutes a persistent external cost. This motivates a finite-time squeezing protocol, where the external coherence is applied only during the initial stage of evolution. Physically, the squeezed reservoir acts as a transient catalyst to accelerate the early buildup of local battery coherence, which in turn enhances the coherent ergotropy. Once this coherence-assisted manifold is established, the squeezed reservoir is no longer required to continuously drive the system.

\begin{figure}[H]
  \centering
  
  \captionsetup{font=small}
  \hspace*{-0.3cm}\includegraphics[width=1.0\linewidth]{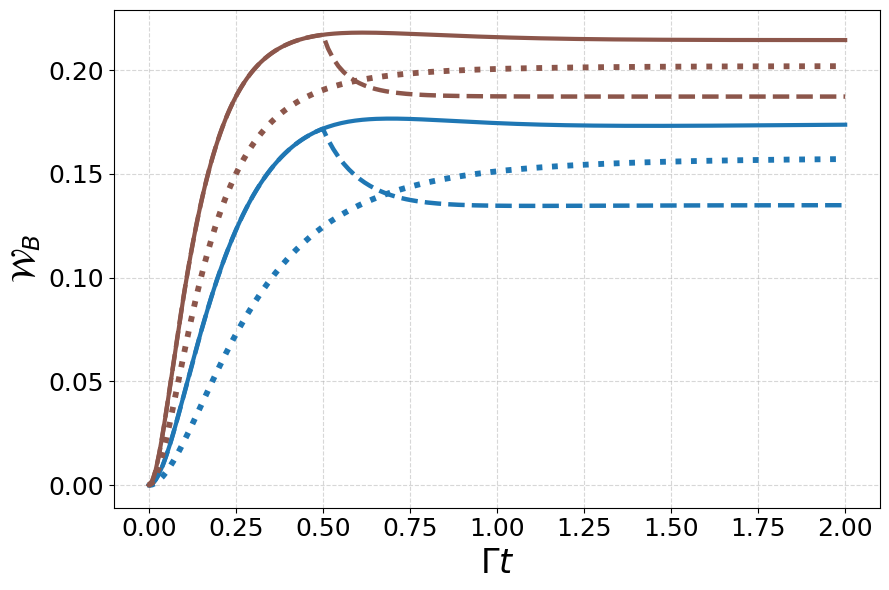}
  \caption{\justifying Time evolution of the battery ergotropy $\mathcal{W}_B$ under three reservoir protocols for two system sizes. Blue curves correspond to $N_B=N_C=4$, and brown curves correspond to $N_B=N_C=8$. Solid lines denote the continuous-squeezing protocol with $r=0.5$ throughout, dashed lines denote the finite-time squeezing protocol in which the squeezing is switched from $r=0.5$ to $r=0$ at $\gamma t=0.5$, and dotted lines denote the vacuum protocol with $r=0$ throughout. Other parameters used are $\theta=\pi/3,~\delta=0$.} 
  \label{fig:quench}
\end{figure}
The finite-time squeezing protocol is implemented via a sudden quench of the squeezing parameter ($r=0.5 \rightarrow r=0$) at a switching time $t_{\text{q}}$. 
This quench approximation is physically justified in the regime $\tau_{\text{sw}} \ll \tau_{\text{rel}} \sim 1/\gamma$, where $\tau_{\text{sw}}$ is the switching duration and $\tau_{\text{rel}}$ denotes the system’s intrinsic relaxation time. Such a timescale hierarchy is achievable in experimental platforms where the control of the input squeezed field is set by large cavity bandwidths \cite{qiu2023broadband}, while the modulation of field intensity and coupling strength remains comparatively slow \cite{bienfait2016controlling}.

 As illustrated in Fig.~\ref{fig:quench}, the finite-time squeezing protocol represents a strategic compromise between the rapid charging speed of the continuous-squeezing protocol and the lower resource cost of the vacuum protocol. During the initial stage, the battery retains the characteristic power boost and accelerated ergotropy growth enabled by the external coherence channel, thereby reaching substantially larger ergotropy at earlier times than in the always-vacuum protocol. After the squeezing is removed, the battery stabilizes at a steady state whose ergotropy remains slightly below that of the vacuum case. In this sense, the finite-time squeezing protocol is resource-efficient: it eliminates the persistent cost of maintaining the squeezed vacuum reservoir while preserving the key transient advantage absent in the vacuum case.

\section{Discussion and Conclusion} \label{sec:conclusion}
In this work, we have developed a framework for quantum coherence-enhanced battery charging in which enhanced battery performance is enabled by coherence generated through two distinct, controllable sources: (i) internal coherence from the charger initial state preparation and (ii) external coherence via reservoir squeezing. In the comparable charger-battery size, resource-efficient regime considered here, initial charger coherence is the key resource for generating and stabilizing large steady-state ergotropy through dark-state protection, whereas its combination with external reservoir squeezing accelerate the transient buildup of ergotropy. The combined action of these two coherence sources determines both the charging rate and the final work-extraction capacity.

By combining analytical results for few-spin systems with numerical simulations for multi-spin configurations, we have identified that the central mechanism underlying both rapid charging and long-time work storage is the buildup of local battery coherence. The charger's polar angle $\theta$ controls the competition between coherent and population-driven contributions to the ergotropy, thereby determining the optimal operating point. In particular, initial charger coherence preparation enables finite and robust steady-state  battery ergotropy even when the available population inversion is limited, while reservoir squeezing further amplifies the early-time charging speed. Both effects are optimized by the phase-matching condition $\delta=0$, which corresponds to optimal alignment between the charger’s initial coherence phase and the reservoir squeezing axis. This coherence-assisted charging preserves the benefits of collective dynamics, including $N^2$ superradiant scaling. Together, these features establish a high-performance regime that surpasses both incoherent and non-collective alternatives.

Our numerical results further distinguish the roles of the two resources in the dynamical evolution. Reservoir squeezing primarily enhances the early-time charging stage by increasing the initial buildup of ergotropy, but this advantage is progressively reduced as dissipation suppresses the squeezing-induced coherence. By contrast, initial charger coherence seeds the dark-state manifold responsible for preserving finite steady-state ergotropy. Consistent with this interpretation, the buildup of charger-battery entanglement tends to reduce the extractable work by increasing the mixedness of the battery’s reduced state.

These results also clarify the role of resource costs. Although a continuously squeezed reservoir can provide a strong transient boost, our dynamical analysis indicates that squeezing need not be maintained throughout the entire evolution. Instead, it can be used during the initial stage as a catalyst for rapidly generating local battery coherence and ergotropy, after which the system can evolve under vacuum dissipation while retaining a substantial fraction of the stored work. This finite-time protocol avoids the continuous cost of maintaining an always-on squeezed reservoir and is therefore both resource-efficient and experimentally realistic. 

Future research will focus on adapting this dissipative framework to generate and stabilize metrologically relevant spin-squeezed states. Such an extension would carry the steady-state protection mechanism identified here beyond quantum batteries and into quantum sensing and related collective-spin platforms.

In summary, these results identify a route toward open quantum batteries that combine enhanced ergotropy, fast charging, and long-time stabilization. By establishing a clear hierarchy of quantum resources, we show that external transient squeezing combined with internal coherent charger preparation provides an optimal balance between charging power, ergotropy stabilization, and experimental feasibility. These findings establish coherence-assisted control as an experimentally viable route toward high-power quantum batteries in platforms ranging from solid-state systems to atomic ensembles.

\section{acknowledgement}
This material is based upon work supported by the Robert A. Welch Foundation (Grant Nos. A-1943-20240404 and A-1261), the Air Force Office of Scientific Research (Award No. FA9550-20-1-0366), the Department of Energy, Fusion Energy Science (Award No. DE-SC0024882, IFE-STAR), and the U.S. Department of Energy, Office of Science, Office of Biological and Environmental Research (Award Number DE-SC-0023103). F. Y. is supported by the Herman F. Heep and Minnie Belle Heep Texas A\&M University Endowed Fund held/administered by the Texas A\&M Foundation.

\section{Data Availability}

The numerical data underlying the figures in this article are available from the corresponding author upon reasonable request.

\appendix
\section{Dynamical equation} \label{appendixa}
The dynamical evolution of the charger-detector spin system's reduced density matrix, $\rho$, is governed by:
\begin{align} \label{dynamical}
\dot{\rho}(t) =&-\frac{i}{\hbar} \mathrm{Tr}_R\left[ \hat{H}_I(t), \rho_s(t_0) \right] \notag \\ 
&- \frac{1}{\hbar^2} \mathrm{Tr}_R \int_{t_0}^{t} dt' \left[ \hat{H}_I(t), \left[ \hat{H}_I(t'), \rho_s(t') \right] \right],
\end{align}
where $\rho_s$ denotes the full system-reservoir composite state. The first term in Eq. (\ref{dynamical}) vanishes because the reservoir expectation values for the displacement operators satisfy  $\langle \hat{a}_\mathbf{k} \rangle = \langle \hat{a}_\mathbf{k}^\dagger \rangle = 0$. Under the assumption of weak system-reservoir coupling, we employ the Born approximation, which allows the total density matrix to be partitioned as $\rho_s(t')\approx \rho(t')\otimes\rho_R$. Here, the reservoir is described by a stationary density operator $\rho_R$ that remains unaffected by the system's dynamics. We further employ the Markovian approximation, where the integrand $\rho_s(t')$ is replaced by $\rho_s(t)$, since dissipation erases the system's memory of its past states. Furthermore, since the system loses memory of its long-past history, it is justified to extend the lower limit of the integral to $-\infty$. These steps yield the quantum master equation:
\begin{align}\label{master1}
\dot{\rho}(t)=-\frac{1}{\hbar^2}\int_{-\infty}^t dt'\mathrm{Tr}_R[H_I(t),[H_I(t'),\rho(t)\otimes\rho_R]].
\end{align}
On inserting the interaction Hamiltonian Eq.~(\ref{Hamiltonian}) into the equation of motion (\ref{master1}), we obtain
\begin{align} \label{master2}
\dot{\rho} = {}&C_0 \left( \hat J^{-} \hat J^{-} \rho + \rho \hat J^{-} \hat J^{-} - 2 \hat J^{-} \rho \hat J^{-} \right) \notag\\ +&  C_0^{*} \left( \hat J^+ \hat J^+ \rho + \rho \hat J^+ \hat J^+ - 2 \hat J^+ \rho \hat J^+ \right) \notag\\
+& C_1 \left( \hat J^{-} \hat J^+ \rho  -  \hat J^+ \rho \hat J^{-} \right) + C_1^* \left( \rho \hat J^{-} \hat J^+   -  \hat J^+ \rho \hat J^{-} \right) \notag\\
+& C_2 \left( \hat J^+ \hat J^{-} \rho  -  \hat J^{-} \rho \hat J^+ \right) +C_2^* \left(  \rho \hat J^+ \hat J^{-} -  \hat J^{-} \rho \hat J^+ \right) ,
\end{align}
where $\hat J^{-}\equiv \hat J^{-}_C+\hat J^{-}_B$ and $\hat J^+\equiv \hat J^+_C+\hat J^+_B$. $C_i$ with $i=0,1,2$ are defined as
\begin{equation}\label{C0}
C_0=- \int_{-\infty}^{t} dt' \sum_{\mathbf{k},\mathbf{k}'} g_\mathbf{k} g_{\mathbf{k}'} 
e^{i(\nu_\mathbf{k} - \omega)t + i(\nu_{\mathbf{k}'} - \omega)t'} 
\langle \hat a_\mathbf{k}^\dagger \hat a_{\mathbf{k}'}^\dagger \rangle,
\end{equation}

\begin{equation} \label{C1}
C_1=- \int_{-\infty}^{t} dt' \sum_{\mathbf{k},\mathbf{k}'} g_\mathbf{k} g_{\mathbf{k}'} 
e^{i(\nu_\mathbf{k} - \omega)t - i(\nu_{\mathbf{k}'} - \omega)t'} 
\langle \hat a_\mathbf{k}^\dagger \hat a_{\mathbf{k}'} \rangle,
\end{equation}

\begin{equation} \label{C2}
C_2=- \int_{-\infty}^{t} dt' \sum_{\mathbf{k},\mathbf{k}'} g_\mathbf{k} g_{\mathbf{k}'} 
e^{-i(\nu_\mathbf{k} - \omega)t + i(\nu_{\mathbf{k}'} - \omega)t'} 
\langle \hat a_\mathbf{k} \hat a_{\mathbf{k}'}^\dagger \rangle.  
\end{equation}
Throughout this work, the reservoir is modeled as a squeezed vacuum or vacuum field. Such fields are characterized as a direct product of single-mode components and thus lack entanglement between different frequency modes. This leads to a Kronecker delta $\delta_{\mathbf{k},\mathbf{k}'}$ in the expectation values appearing in the integrands of Eqs.~(\ref{C0}),~(\ref{C1}), and~(\ref{C2}), enforcing the condition $\mathbf{k} = \mathbf{k}'$. Furthermore, in a 3D cavity, the sum over $\mathbf{k}$ may be replaced by an integral through the prescription $\sum_\mathbf{k} \rightarrow V_\mathrm{ph}/{\pi^2} \int_0^\infty dk \, k^2$, where $V_\mathrm{ph}$ is the photon volume. By applying the identity $\int_{-\infty}^{t} dt' e^{i(\omega - \nu_\mathbf{k})t'} = \pi \delta(\omega - \nu_\mathbf{k})$, the expressions for $C_i$ can be further simplified to

\begin{equation}
C_0=-\frac{V_\mathrm{ph}}{\pi}|\mathbf{k}_0|^2 g_{\mathbf{k}_0}^2  \langle \hat a_{\mathbf{k}_0}^\dagger \hat a_{\mathbf{k}_0}^\dagger \rangle,
\end{equation}

\begin{equation}
C_1=-\frac{V_\mathrm{ph}}{\pi}|\mathbf{k}_0|^2 g_{\mathbf{k}_0}^2 \langle \hat a^\dagger_{\mathbf{k}_0} \hat a_{\mathbf{k}_0} \rangle,
\end{equation}

\begin{equation}
C_2=-\frac{V_\mathrm{ph}}{\pi}|\mathbf{k}_0|^2 g_{\mathbf{k}_0}^2 \langle \hat a_{\mathbf{k}_0} \hat a_{\mathbf{k}_0}^\dagger \rangle.
\end{equation}
Here, $\mathbf{k}_0$ satisfies the resonance condition $\nu_{\mathbf{k}_0} = \omega$. Physically, this reflects the fact that only the $\mathbf{k}_0$th photon mode resonant with the spin frequency interacts with the spins. Therefore, in the following analysis of the squeezed vacuum and squeezed thermal reservoir cases, it suffices to consider only the mode $\mathbf{k}_0$ in the reservoir. Obviously, both $C_1$ and $C_2$ are real numbers. Defining $\gamma={V_\mathrm{ph}}|\mathbf{k}_0|^2 g_{\mathbf{k}_0}^2/\pi$, $\bar{n}=\langle a^\dagger_{\mathbf{k}_0} \hat a_{\mathbf{k}_0} \rangle$, and $\bar{m}=\langle \hat a_{\mathbf{k}_0} \hat a_{\mathbf{k}_0} \rangle$, Eq.~(\ref{master2}) is simplified to 
\begin{align} \label{master3a}
\dot{\rho} ={} & \gamma (\bar{n}+1)~\mathcal{L}[\hat J^{-}] \rho + \gamma \bar{n}~\mathcal{L}[\hat J^+] \rho \notag\\+ & \gamma\bar{m}^* \left( 2 \hat J^{-} \rho \hat J^{-} -\hat J^{-} \hat J^{-} \rho - \rho \hat J^{-} \hat J^{-} \right) \notag\\
+ &\gamma\bar{m} \left( 2 \hat J^+ \rho \hat J^+ -\hat J^+ \hat J^+ \rho - \rho \hat J^+ \hat J^+  \right)  .
\end{align}
$\mathcal{L}$ is the Lindblad operator, defined by
\begin{align} 
\mathcal{L}[\hat O] \, \rho = 2 \hat O \rho \hat O^\dagger - \hat O^\dagger \hat O \rho - \rho \hat O^\dagger \hat O.
\end{align}

\section{State Representations and Symmetry Transformations}\label{appendixb}
\subsection{Dicke Basis Expansion of the Charger State}
The charger is initialized in the collective spin-coherent state $|\theta,\phi\rangle_C$, defined by the rotation $|\theta,\phi\rangle_C = \hat{R}_z(\phi)\hat{R}_y(\theta)|J,J\rangle$. Expanding this state in the Dicke basis $|J,M\rangle$ yields:
\begin{equation}
  \ket{\theta,\phi}_C
  = \sum_{M=-J}^{J} C_M(\theta,\phi)\,\ket{J,M},
\end{equation}
where the coefficients are given by the generalized binomial distribution:

\begin{align}
  C_M(\theta,\phi)= \sqrt{\binom{2J}{J+M}}
    \left(\cos\frac{\theta}{2}\right)^{J+M}
    \left(\sin\frac{\theta}{2}\right)^{J-M}
    e^{-iM\phi}.
\end{align}
This expansion makes explicit that the spin-coherent state is a coherent superposition of Dicke states with different magnetic quantum numbers $M$. The polar angle $\theta$ controls the population weights across the Dicke ladder. Specifically, the `top' of this ladder ($M=J$) corresponds to the fully excited state $|J,J\rangle_C$ defined in Eq.~(\ref{spco}), while varying $\theta$ redistributes population among the lower $M$ sectors to inject coherence into the system. Simultaneously, the azimuthal angle $\phi$ imprints well-defined relative phases between these different $M$ sectors.

\subsection{Dependence on Relative Phase}
\label{app:phase_reduction}

The system dynamics are governed by a single combined phase parameter $\delta$, rather than by the charger phase $\phi$ and reservoir phase $\varphi$ independently. As established in Sec.~\ref{sec:system}, the evolution is governed by the master equation $\dot{\rho} = 2 \hat L_{\varphi} \rho \hat L_{\varphi}^\dagger - \hat L_{\varphi}^\dagger \hat L_{\varphi} \rho - \rho \hat L_{\varphi}^\dagger \hat L_{\varphi}$. Crucially, the jump operator defined in Eq.~(\ref{eq:jump_op}) carries the reservoir phase information:
\begin{equation}
\hat L_{\varphi} = \hat J^{-} \cosh(r) e^{ \frac{i\varphi}{2}} + \hat J^+ \sinh(r)e^{-\frac{i \varphi}{2}}. 
\label{eq:b2}
\end{equation}

To demonstrate the reduction to a single phase, we consider the charger initialized in the collective spin-coherent state $|\theta,\phi\rangle_C$. We transform to a rotating reference frame defined by the unitary operator $\hat{R}_z(-\phi) = e^{i\phi \hat{J}_C^z}$. Under this rotation, the collective ladder operators transform as $\hat R_z(-\phi) \hat J^\pm \hat R_z^\dagger(-\phi) = e^{\pm i\phi} \hat J^\pm$. Consequently, the jump operator in Eq.~(\ref{eq:b2}) transforms as
\begin{align}
\hat R_z(-\phi) \hat L_\varphi \hat R_z^\dagger(-\phi) = \hat L_{\varphi-2\phi}.
\label{eq:Lshift}
\end{align}
In this rotated frame, the density matrix $\rho' = \hat R_z(-\phi)\rho\hat R_z^\dagger(-\phi)$ evolves according to
\begin{align}
\dot\rho' \;=\; \gamma\hat R_z(-\phi)\,\mathcal{L}_\varphi\!\big[\hat R_z^\dagger(-\phi)\rho' \hat R_z(-\phi)\big]\,\hat R_z^\dagger(-\phi).
\label{eq:covariance}
\end{align}
Substituting Eq.~(\ref{eq:Lshift}) into Eq.~(\ref{eq:covariance}), the master equation reduces to
\begin{align}
\dot\rho' \;=\; \gamma\mathcal{L}_{\varphi-2\phi}[\rho'].
\label{eq:rotatedrho}
\end{align}
Crucially, the initial spin-coherent state transforms as $\ket{\theta,\phi}_C\to \ket{\theta,0}_C$.
This implies that the dynamics of a system in the rotated reference frame with parameters $(\theta, \phi, \varphi)$ are mathematically equivalent to a system with parameters $(\theta, 0, \varphi-2\phi)$. Therefore, all physical observables, such as the stored energy and ergotropy, are functions of the combined phase $\delta$.

The physical origin of this $\delta$-dependence becomes intuitive when considering that the master equation is second-order in the interaction Hamiltonian $H_I$, as shown in Eq.~(\ref{master1}). When the reservoir is traced out, the phase-carrying contributions from the squeezed bath arise solely from its anomalous two-photon correlations, $\hat{a} \hat{a} \propto e^{-i\varphi}$. These correlations naturally pair with two collective spin-raising operators to produce terms of the form $\hat{J}^{+} \hat{J}^{+}$ (and similarly $\hat{a}^\dagger \hat{a}^\dagger $ with $\hat{J}^{-} \hat{J}^{-}$), as shown in Eq.~(\ref{master2}). Since the operator product $\hat{J}^{+} \hat{J}^{+}$ carries the spin azimuthal phase $e^{2i\phi}$, the contributions of the effective coupling terms, such as $\hat{a}\hat{J}^+$ and $\hat{a}^\dagger\hat{J}^-$, depend exclusively on the combined factor $e^{\pm i(\varphi - 2\phi)/2}$. Consequently, the charging dynamics are controlled by the relative phase $\delta = \varphi - 2\phi$, rather than by the charger phase $\phi$ and squeezing phase $\varphi$ independently.

Finally, for the special case $\theta = 0$, both the charger and battery domains are initially polarized along the $z$ axis. In this case, the azimuthal angle $\phi$ contributes only a global phase to the initial charger state,
\begin{align}
\ket{\theta=0,~\phi}_C=e^{-i\phi N_C/2}\ket{\uparrow\cdots\uparrow}_C.
\label{eq:iniphi}
\end{align}
which has no effect on physical observables. Since all observables depend only on the combination $\delta$, the irrelevance of $\phi$ immediately implies that they are also independent of the squeezing phase $\varphi$. Consequently, for $\theta=0$, quantities such as the stored energy and ergotropy are completely independent of both $\phi$ and $\varphi$.

\section{Few-spin system analytical steady-state solutions and collective advantage}
\label{appendixc}
In this appendix, we assume the symmetric spin ratio $N_C = N_B = N$. 

\textbf{Minimal Two-Spin System ($N=1$):}
We begin by examining the analytical steady-state solution for the minimal case of $N= 1$. The steady-state manifold is spanned by the dark states $|\Psi_{d_i}\rangle$, which are defined by the condition $\hat{L}_\varphi |\Psi_{d_i}\rangle = 0$. This condition yields two linearly independent solutions that remain protected from environmental dissipation:
\begin{align}
\ket{\Psi_{d_1}} &= \ket{\psi_-} = \frac{\ket{\uparrow\downarrow}-\ket{\downarrow\uparrow}}{\sqrt{2}}, \label{dark1}\\[4pt]
\ket{\Psi_{d_2}} &= \frac{\sqrt{1+\bar{n}}}{\sqrt{1+2\bar{n}}}\left( \frac{|\bar{m}|e^{i\varphi}}{1+\bar{n}}\ket{\uparrow\uparrow} - \ket{\downarrow\downarrow} \right).
\label{dark2}
\end{align}
Here, $\ket{\Psi_{d_1}}$ is the singlet state, fully protected under dissipation, while $\ket{\Psi_{d_2}}$ interpolates between $\ket{\downarrow\downarrow}$ and an entangled superposition depending on the squeezing strength. The corresponding non-decaying operators are the projectors and cross terms:
\begin{equation}
|\Psi_{d_1}\rangle\langle\Psi_{d_1}|,~ |\Psi_{d_2}\rangle\langle\Psi_{d_2}|,~
|\Psi_{d_2}\rangle\langle\Psi_{d_1}|,~ |\Psi_{d_1}\rangle\langle\Psi_{d_2}|,
\label{darkmat}
\end{equation}
which remain invariant under the dynamical evolution. 

While the reservoir engineering defines the available dark-state manifold, as shown in Eqs.~(\ref{dark1}) and~(\ref{dark2}), the specific weights of the stationary operators in Eq.~(\ref{darkmat}) are determined by the charger's initial state. Consider the single-spin charger prepared in the spin-coherent state $\ket{\theta,\phi}_{C} =\cos(\theta/2) ~\ket{\uparrow}_C + e^{i\phi}\sin(\theta/2) ~\ket{\downarrow}_C$, the analytical steady state evolving from the initial system state $\ket{\theta,\phi}_{C}\ket{\downarrow}_B$ in the case $N = 1$ can be obtained:

\begin{widetext}

\begin{align}\label{rhoss_co1}
\rho_{ss} &= \frac{1+\cos\theta}{4}\,|\Psi_{d_1}\rangle\langle\Psi_{d_1}|
         +  \frac{3-\cos\theta}{4}\,|\Psi_{d_2}\rangle\langle\Psi_{d_2}|  - \frac{e^{i\phi}\sin\theta}{2}\frac{\sqrt{1+\bar n}}{\sqrt{2+4\bar n}}
        |\Psi_{d_2}\rangle\langle\Psi_{d_1}|
        -  \frac{e^{-i\phi}\sin\theta}{2}\frac{\sqrt{1+\bar n}}{\sqrt{2+4\bar n}}
            |\Psi_{d_1}\rangle\langle\Psi_{d_2}|.
\end{align}

The reduced density matrix of the battery, $\rho_B$, is obtained by evaluating the partial trace over the charger subsystem for each term in Eq.~(\ref{rhoss_co1}):
\begin{align}
\operatorname{Tr}_C\!\left[\,|\Psi_{d_1}\rangle\langle\Psi_{d_1}|\,\right]
&= \tfrac12\Big( |\uparrow\rangle\langle\uparrow| 
                 + |\downarrow\rangle\langle\downarrow| \Big),
\label{D2}
\\[8pt]
\operatorname{Tr}_C\!\left[\,|\Psi_{d_2}\rangle\langle\Psi_{d_2}|\,\right]
&= 
\frac{|\bar m|^2}{(1+\bar n)(1+2\bar n)}\,|\uparrow\rangle\langle\uparrow|
+
\frac{1+\bar n}{1+2\bar n}\,|\downarrow\rangle\langle\downarrow|,
\label{D3}
\\[8pt]
\operatorname{Tr}_C\!\left[\,|\Psi_{d_2}\rangle\langle\Psi_{d_1}|\,\right]
&=
\frac{1}{\sqrt{2}}
\left(
\frac{\sqrt{1+\bar n}}{\sqrt{1+2\bar n}}
         \frac{|\bar m|e^{i\varphi}}{1+\bar n}|\uparrow\rangle\langle\downarrow|
+
\frac{\sqrt{1+\bar n}}{\sqrt{1+2\bar n}}|\downarrow\rangle\langle\uparrow|
\right),
\label{D4}
\\[8pt]
\operatorname{Tr}_C\!\left[\,|\Psi_{d_1}\rangle\langle\Psi_{d_2}|\,\right]
&=
\frac{1}{\sqrt{2}}
\left(
\frac{\sqrt{1+\bar n}}{\sqrt{1+2\bar n}}
         \frac{|\bar m|e^{-i\varphi}}{1+\bar n}|\downarrow\rangle\langle\uparrow|+
\frac{\sqrt{1+\bar n}}{\sqrt{1+2\bar n}}|\uparrow\rangle\langle\downarrow|
\right),
\label{D5}
\end{align}

Evaluating the partial traces of the stationary operators in Eqs.~(\ref{D2})-(\ref{D5}) yields the steady-state battery density matrix. In the energy eigenbasis $\{|\downarrow\rangle, |\uparrow\rangle\}$, this reduced state is given by:

\begin{equation}
\rho_B = \frac{1}{8(1+2\bar{n})}
\begin{pmatrix}
7 + 8\bar{n} - \cos\theta & -2\sin\theta \left(1+ \bar{n}  + |\bar{m}|e^{-i\delta} \right) \\
-2\sin\theta \left( 1+\bar{n}  + |\bar{m}|e^{i\delta} \right) & 1 + 8\bar{n} + \cos\theta 
\end{pmatrix}.
\label{densi_co1}
\end{equation}

While Eq. (\ref{densi_co1}) (corresponding to Eq. (\ref{densi_co}) in the main text) contains explicit phase factors $e^{\pm i\delta}$, the eigenvalues of $\rho_B$ depend only on the relative phase $\delta$. Since the ergotropy is determined uniquely by the spectrum, the global phase $\delta$ does not influence the extractable work, justifying its omission in the main text expression. Physically, the diagonal populations in $\rho_B$ originate from the projectors $|\Psi_{d_1}\rangle\langle\Psi_{d_1}|$ and $|\Psi_{d_2}\rangle\langle\Psi_{d_2}|$, whereas the off-diagonal elements—solely responsible for the emergence of battery coherence—arise from the cross terms $|\Psi_{d_2}\rangle\langle\Psi_{d_1}|$ and $|\Psi_{d_1}\rangle\langle\Psi_{d_2}|$. The battery ergotropy is found to be:
\begin{align}\label{ergo_co}
W_B = \frac{-3 + \cos\theta + \sqrt{%
10 - 6\cos\theta + (3+12\bar n +8\bar n^2)\sin^2\theta
+ 8(1+\bar n)|\bar m|\sin^2\theta \cos \delta%
}}{8(1+2\bar n)} .
\end{align}
To prove that $W_B \ge 0$ across the entire parameter space, we rationalize the numerator of Eq.~(\ref{ergo_co}) to obtain the equivalent form:
\begin{align}\label{ergo_co>0}
W_B&= \frac{-3 + \cos\theta + \sqrt{%
10 - 6\cos\theta + (3+12\bar n +8\bar n^2)\sin^2\theta
+ 8(1+\bar n)\lvert\bar m\rvert \sin^2\theta \cos \delta}}{8(1+2\bar n)} \notag \\
&= \frac{4\left(1+2\bar n+2\lvert\bar m\rvert \cos \delta\right)(1+\bar n)\sin^2\theta}{%
8(1+2\bar n)\left[3 - \cos\theta + \sqrt{%
10 - 6\cos\theta + (3+12\bar n +8\bar n^2)\sin^2\theta
+ 8(1+\bar n)\lvert\bar m\rvert \sin^2\theta \cos \delta}\right]}.
\end{align}
The non-negativity of $W_B$ is guaranteed by the structure of the numerator, specifically by the bracketed term:
\begin{align}
1 +2\bar n  + 2|\bar m|\cos\delta
\;\ge\; 1 +2\bar n - 2|\bar m|
\;=\; 1 +2\bar n - 2\sqrt{\bar n(1+\bar n)} \;>\; 0 .
\end{align}
This result confirms that the extractable work remains non-negative across the entire parameter space, vanishing only in the limit $\theta = 0$. As shown in the steady-state battery density matrix $\rho_B$, this limit corresponds precisely to the absence of initial charger coherence, reinforcing the role of internal coherence as the primary driver of ergotropy extraction in the single-spin limit.

\textbf{Multi-Spin Ensembles ($N=2$):}
Next, we demonstrate that for multi-spin ensembles, the battery can acquire nonzero ergotropy even without initial charger coherence ($\theta=0$). Specifically, for the case $N = 2$, we assume the charger is initialized in the fully excited state ($\theta=\phi=0$). Following the derivation in Appendix~\ref{appendixd}, the steady-state reduced density matrix of the battery in the symmetric triplet basis $\{|J_B = 1, M_B = -1, 0, 1\rangle\}$ is:

\begin{align}
\rho_B = \frac{1}{12(1+2\bar{n})(3+8\bar{n}+8\bar{n}^2)}
\scalebox{0.95}{$
\begin{pmatrix}
19 + 79\bar{n} + 120\bar{n}^2 + 64\bar{n}^3 & 0 & 3|\bar{m}|e^{-i\delta} \\
0 & 13 + 58\bar{n} + 96\bar{n}^2 + 64\bar{n}^3 & 0 \\
3|\bar{m}|e^{i\delta} & 0 & 4 + 31\bar{n} + 72\bar{n}^2 + 64\bar{n}^3
\end{pmatrix}$}.
\label{rhoss4}
\end{align}

\end{widetext}
resulting in zero battery state coherence.
Although an analytical expression for the ergotropy can be obtained for the N=2 case, we omit it here because it is prohibitively lengthy and adds little physical insight beyond the density matrix in Eq. (\ref{rhoss4}). Nevertheless, the $N=2$ result in Eq. (\ref{rhoss4}) reveals a qualitative shift: unlike the $N=1$ case [Eq.~(\ref{densi_co1})], where off-diagonal elements vanish at $\theta = 0$, the multi-spin architecture maintains nonzero coherence even in this limit. This qualitative shift occurs because the initial state $|\uparrow\uparrow\rangle_C|\downarrow\downarrow\rangle_B$ possesses a non-vanishing overlap with the dark-state cross terms $|\Psi_{J,0}\rangle\langle\Psi_{K,0}|$. Such an interaction is structurally impossible in the single-spin limit ($N=1$), as the initial system state $|\uparrow\rangle_C|\downarrow\rangle_B$ has zero overlap with the available cross-term operators [Eqs.~(\ref{D2})--(\ref{D5})], resulting in vanishing battery coherence. In the general multi-spin case, the number of these dark-state cross terms increases proportionally to $N^2$, as detailed in Appendix~\ref{appendixd}, providing the foundational mechanism for the enhanced ergotropy storage analyzed in the main text.

As explained in Sec.~\ref{sec:ergo_cohe}, QB ergotropy originates from two distinct physical mechanisms: population inversion among the energy eigenstates, quantified by the incoherent contribution $W_B^P$, and quantum coherence between these states, which enables work extraction even in the absence of inversion, quantified by $W_B^C$. Under collective coupling to a common vacuum reservoir, the initial state within each Dicke subspace relaxes to its respective dark (subradiant) state that is immune to the dominant dissipation channels. Crucially, moving to a multi-spin architecture provides two complementary paths to ergotropy enhancement. First, the increased number of dark-state cross-terms facilitates a much larger buildup of local battery coherence, thereby significantly increasing the coherent ergotropy $W_B^C$. Simultaneously, the richer level structure structurally enables population inversions ($W_B^P$) across multiple energy levels that are otherwise forbidden in the single-spin limit. By comparing the emergence of these two ergotropy components at $N=1$ and $N=2$, we establish the foundational mechanism driving the enhanced ergotropy storage and superextensive scaling analyzed numerically in the main text; these few-spin results thus provide the prototypical proof for the collective advantages observed in larger ensembles.

\section{General solutions of Steady State of batteries with multiple spins} \label{appendixd}
This appendix outlines the numerical procedure for obtaining steady-state solutions of the master equation [Eq. (\ref{dynamical2})] for multi-spin quantum batteries (QBs) under a squeezed reservoir. The dynamics of the density matrix $\rho$ are described by the Liouvillian superoperator $\mathcal{L}$—a linear operator that maps the state to its time derivative, $\dot{\rho} = \mathcal{L}\rho$, accounting for both the unitary evolution and the dissipative environment through the jump operator $\hat{L}_{\varphi}$  [Eq. (\ref{eq:jump_op})]:

\begin{equation}
\dot{\rho} = \gamma\left(2 \hat L_{\varphi} \rho \hat L_{\varphi}^\dagger - \hat L_{\varphi}^\dagger \hat L_{\varphi} \rho - \rho \hat L_{\varphi}^\dagger \hat L_{\varphi}\right).
\label{master4}
\end{equation}
The efficiency of this numerical analysis stems from the structure of $\hat{L}_{\varphi}$. Because it is composed of collective spin operators $\hat J^{\pm} = \hat J_C^{\pm} + \hat J_B^{\pm}$, the total angular momentum is conserved throughout the dynamics. This symmetry permits a block-diagonal decomposition of the Liouvillian in the Dicke state basis $\{|J,M\rangle\}$, which we adopt as our primary basis throughout this appendix.

In our QB setup, the charger-battery system is initialized in the product state $|\theta,\phi\rangle_C \otimes |\downarrow\downarrow\cdots\downarrow\rangle_B$. This state restricts the system to the symmetric manifolds $J_C = N_C/2$ and $J_B = N_B/2$, confining the dynamics to invariant $J$-sectors ranging from $J_{\mathrm{min}} = |J_C - J_B|$ to $J_{\mathrm{max}}=J_C + J_B$. Within each independent sector—represented as a Dicke ladder spanned by $M = -J, \dots, J$—the task reduces to finding the zero-eigenvalue solution (the dark state) of the Liouvillian. Because $\hat L_{\varphi}$ is non-Hermitian, these stationary solutions are most naturally expressed using a biorthogonal set of right and left eigenstates:
\[
\hat L_{\varphi}|\Psi_{J,M}\rangle = \lambda_{J,M} |\Psi_{J,M}\rangle, \qquad
\hat L_{\varphi}^\dagger |\Phi_{J,M}\rangle = \lambda_{J,M}^* |\Phi_{J,M}\rangle,
\]
where $\{\ket{\Psi_{J,M}}\}$ and $\{\ket{\Phi_{J,M}}\}$ form a biorthogonal pair satisfying $\langle \Phi_{J,M} | \Psi_{J',M'} \rangle = \delta_{J,J'}\delta_{M,M'}$.
The eigenvalues are $\lambda_{J,M} = \sqrt{2 \sinh(2r(\omega))}\, M$. The explicit forms of these eigenstates, first derived in
Refs.~\cite{aragone1974intelligent,rashid1978intelligent}, are
\begin{align}\label{eigenR}
|\Psi_{J,M}\rangle &= A\, 
\exp\left[\left(\alpha + \frac{i\varphi}{2}\right)\hat J^z\right]\,
\exp\!\left(-\frac{i\pi}{2}\hat J^y\right)|J,M\rangle, \notag \\
|\Phi_{J,M}\rangle &= B\,
\exp\left[\left(-\alpha + \frac{i\varphi}{2} \right) \hat  J^z\right]\,
\exp\!\left(-\frac{i\pi}{2}\hat J^y\right)|J,M\rangle,
\end{align}
where the constants $A$ and $B$ are chosen to satisfy $\langle \Phi_{J,M} | \Psi_{J,M} \rangle = 1$. The parameter $\alpha$ is related to the squeezing strength through $\exp(2\alpha) = \tanh(2r)$, where $r$ is the squeezing parameter. These biorthogonal states provide a natural basis for analyzing the action of the squeezed-reservoir dissipator. Since $\hat L_{\varphi}$ preserves total spin, completeness holds independently within each Dicke sector:
\begin{align}\label{normalization}
\sum_{M=-J}^{J} |\Phi_{J,M}\rangle\langle\Psi_{J,M}| = \mathbb{I}_J,
\end{align}
where $\mathbb{I}_J$ projects onto the Dicke subspace with fixed total spin $J$.

The structure of the dark states and the resulting steady-state manifold depends on whether the total number of spins is even or odd. For an \textbf{even total number of spins}, both the total spin $J$ and the magnetic quantum number $M$ are integers. 
Each Dicke ladder therefore contains a state with $M = 0$ that satisfies $\hat L_{\varphi}|\Psi_{J,0}\rangle = 0$ and $\hat L_{\varphi}^\dagger|\Phi_{J,0}\rangle = 0$. Consequently, the projector $|\Psi_{J,0}\rangle\langle\Psi_{J,0}|$ is annihilated by the dissipative part of the master equation and remains invariant under the evolution. While the left eigenvectors $|\Phi_{J,0}\rangle$ are required for the biorthogonal decomposition, they do not represent physical stationary configurations. The right eigenvectors $|\Psi_{J,0}\rangle$ are therefore identified as the \emph{dark states} of the dynamics.

Since each distinct value of total spin $J$ contributes exactly one such dark state, the total number of dark states equals the number of ladders, $N_J=\mathrm{min}(N_C, N_B)+1$. Moreover, the block-diagonal nature of the Liouvillian ensures that not only are the projectors $|\Psi_{J,0}\rangle\langle\Psi_{J,0}|$ invariant, but the cross-terms $|\Psi_{J,0}\rangle \langle\Psi_{K,0}|$ (where $J \neq K$) are also stationary. Therefore, the most general steady state in the even-spin case is an arbitrary linear combination of these operators:
\begin{align}
\rho_{\mathrm{ss}}^{\mathrm{even}} = \sum_{J,K=J_{\mathrm{min}}}^{J_{\mathrm{max}}} C_{JK} |\Psi_{J,0}\rangle \langle \Psi_{K,0}|,
\label{rho_even}
\end{align}
where the coefficients satisfy the Hermiticity condition $C_{JK} = C_{KJ}^*$. To determine $C_{JK}$ for a given initial state, we utilize the Liouville-space representation of the master equation~\eqref{master4}. By mapping the density matrix to its column-vectorized form $\vec{\rho}$, the dynamics are governed by the vectorized Liouvillian superoperator $\mathcal{L}_s$ :
\begin{equation}\label{LvR}
\dot{\vec\rho}=\gamma\mathcal{L}_s\vec\rho,
\end{equation}
where
\begin{equation}
\mathcal{L}_s
= 2\, \hat L_{\varphi}^* \otimes \hat L_{\varphi}
 - \mathbb{I} \otimes (\hat L_{\varphi}^\dagger \hat L_{\varphi})
 - (\hat L_{\varphi}^{T} \hat L_{\varphi}^*) \otimes \mathbb{I}
 \label{master5}
\end{equation}
represents the dissipative evolution in the coupled basis. Since the charger-battery system is spanned by $D=(N_C+1)(N_B+1)$ Dicke basis states, the matrices $\mathbb{I}$ and $\hat L_{\varphi}$ have dimension $D \times D$. Consequently, the superoperator $\mathcal{L}_s$ and the vectorized state $\vec{\rho}$ possess dimensions $D^2 \times D^2$ and $D^2$, respectively.

The stationary states of the vectorized master equation correspond to the right null-space vectors of $\mathcal{L}_s$. Since the set of operators $\{|\Psi_{J,0}\rangle\langle\Psi_{K,0}|\}$ forms a complete basis for the steady-state manifold, we define their vectorized counterparts $\vec{v}_{JK}$ as the basis for this null space. These vectorized basis elements are explicitly expressed as:
\begin{align}\label{nullv}
\vec{v}_{JK}
= \vec{\Psi}_{K,0}^{\,*} \otimes \vec{\Psi}_{J,0},
\end{align}
where $\vec{\Psi}_{J,0}$ denotes the column-vector representation of the dark state $|\Psi_{J,0}\rangle$ in the Dicke basis. Because these dark states satisfy the condition $\hat L_{\varphi}\,\vec{\Psi}_{J,0}=0$ , one can verify directly that $\vec v_{JK}$ is a right null-vector of the Liouvillian  by substituting 
Eq.~(\ref{nullv}) into Eq.~(\ref{LvR}) and applying the mixed-product property of the Kronecker product:
\[
(A \otimes B)(\vec{a} \otimes \vec{b}) = (A\vec{a}) \otimes (B\vec{b}).
\]
To construct the final steady state, we also introduce the corresponding left null-space eigenvectors $\vec w_{JK}$, defined by $\mathcal{L}_s^\dagger \vec w_{JK}=0$ and the biorthogonality condition $\vec w_{JK}^\dagger \cdot \vec v_{J'K'} = \delta_{JJ'}\delta_{KK'}$. With these dual bases, the vectorized steady state can be written as the projection of the vectorized initial state $\vec \rho_0$ onto the Liouvillian null space:
\begin{equation}\label{vecrho}
\vec{\rho}_{\mathrm{ss}}^{\,\,\mathrm{even}}=\sum_{J,K=J_{\mathrm{min}}}^{J_{\mathrm{max}}}   (\vec{w}_{JK}^\dagger\cdot\vec{\rho}_0) \vec{v}_{JK}.
\end{equation}

The coefficients $C_{JK} = \vec{w}_{JK}^\dagger \cdot \vec{\rho}_0$ represent the stationary weights determined by the initial preparation. In particular, the diagonal terms $C_{JJ}$ quantify the total population trapped within each invariant $J$-sector. Because the Liouvillian preserves the trace within each independent block, the subspace projector $\mathbb{I}_J$ serves as a conserved quantity and corresponds to the left-null vector $\vec w_{JJ} = \mathrm{vec}(\mathbb{I}_J)$. The associated coefficient in Eq.~(\ref{vecrho}) is then given by:

\begin{equation}\label{coeff2}
C_{JJ}= (\vec{w}_{JJ}^\dagger\cdot\vec{\rho}_0)=\mathrm{Tr}(\rho_0\cdot\mathbb{I}_J)=\sum_{M=-J}^{J}\langle J,M|\rho_0|J,M\rangle .
\end{equation}
Physically, this indicates that the entire population initially residing in a given $J$-ladder relaxes into the corresponding dark state  $|\Psi_{J,0}\rangle\langle\Psi_{J,0}|$.

The distinction in the nature of the stationary states between even and odd total spin arises from the available magnetic quantum numbers $M$ within each Dicke ladder. For an \textbf{even total number of spins}, the total spin $J$ takes integer values, and every ladder contains an $M=0$ state. Because the jump operator satisfies $\hat L_\varphi |J,0\rangle = 0$, each $J$-sector admits a pure dark state $|\Psi_{J,0}\rangle$, leading to the steady-state manifold described in Eq.~(\ref{rho_even}). Conversely, for an \textbf{odd total number of spins}, all $J$ values are half-integers and no ladder contains an $M=0$ state. Consequently, no pure state within these $J$-sectors satisfies $\hat L_\varphi|\psi\rangle=0$, and the stationary states are necessarily mixed \cite{agarwal1990cooperative}. These states are restricted to the diagonal $J=K$ blocks, where the steady state $\rho_J$ of each sector takes the form:
\begin{align}
\rho_{J} = \tilde{A} \sum_{M,N=-J}^{J}  \frac{\langle \Phi_{J,M} | \Phi_{J,N} \rangle}{M N} \, |\Psi_{J,M}\rangle \langle \Psi_{J,N}|,
\label{odddark}
\end{align}
where $\tilde{A}$ is a normalization constant. Since the Liouvillian still preserves the trace within each independent $J$-block, the stationary weights $C_{JJ}$ are determined by the same projection logic as the even-spin case [Eq. (\ref{coeff2})]. The total steady state for an odd number of spins is therefore:
\begin{align}
\rho_{\mathrm{ss}}^{\mathrm{odd}} = \sum_{J=J_{\mathrm{min}}}^{J_{\mathrm{max}}} C_{JJ} \, \rho_J.
\label{oddss}
\end{align}
The absence of $J \neq K$ stationary cross terms in the odd-spin case implies the absence of the cross-ladder interference responsible for generating local battery coherence. Consequently, the steady-state coherent ergotropy is strongly suppressed when the total number of spins is odd. To focus on the regime where dark-state protection most effectively stabilizes these coherence-driven advantages, the main text primarily analyzes systems with an even total number of spins.

\bibliography{QBSqueezedref.bib}

\begin{thebibliography}{51}%
\makeatletter
\providecommand \@ifxundefined [1]{%
 \@ifx{#1\undefined}
}%
\providecommand \@ifnum [1]{%
 \ifnum #1\expandafter \@firstoftwo
 \else \expandafter \@secondoftwo
 \fi
}%
\providecommand \@ifx [1]{%
 \ifx #1\expandafter \@firstoftwo
 \else \expandafter \@secondoftwo
 \fi
}%
\providecommand \natexlab [1]{#1}%
\providecommand \enquote  [1]{``#1''}%
\providecommand \bibnamefont  [1]{#1}%
\providecommand \bibfnamefont [1]{#1}%
\providecommand \citenamefont [1]{#1}%
\providecommand \href@noop [0]{\@secondoftwo}%
\providecommand \href [0]{\begingroup \@sanitize@url \@href}%
\providecommand \@href[1]{\@@startlink{#1}\@@href}%
\providecommand \@@href[1]{\endgroup#1\@@endlink}%
\providecommand \@sanitize@url [0]{\catcode `\\12\catcode `\$12\catcode `\&12\catcode `\#12\catcode `\^12\catcode `\_12\catcode `\%12\relax}%
\providecommand \@@startlink[1]{}%
\providecommand \@@endlink[0]{}%
\providecommand \url  [0]{\begingroup\@sanitize@url \@url }%
\providecommand \@url [1]{\endgroup\@href {#1}{\urlprefix }}%
\providecommand \urlprefix  [0]{URL }%
\providecommand \Eprint [0]{\href }%
\providecommand \doibase [0]{https://doi.org/}%
\providecommand \selectlanguage [0]{\@gobble}%
\providecommand \bibinfo  [0]{\@secondoftwo}%
\providecommand \bibfield  [0]{\@secondoftwo}%
\providecommand \translation [1]{[#1]}%
\providecommand \BibitemOpen [0]{}%
\providecommand \bibitemStop [0]{}%
\providecommand \bibitemNoStop [0]{.\EOS\space}%
\providecommand \EOS [0]{\spacefactor3000\relax}%
\providecommand \BibitemShut  [1]{\csname bibitem#1\endcsname}%
\let\auto@bib@innerbib\@empty
\bibitem [{\citenamefont {Scully}\ \emph {et~al.}(2003)\citenamefont {Scully}, \citenamefont {Zubairy}, \citenamefont {Agarwal},\ and\ \citenamefont {Walther}}]{doi:10.1126/science.1078955}%
  \BibitemOpen
  \bibfield  {author} {\bibinfo {author} {\bibfnamefont {M.~O.}\ \bibnamefont {Scully}}, \bibinfo {author} {\bibfnamefont {M.~S.}\ \bibnamefont {Zubairy}}, \bibinfo {author} {\bibfnamefont {G.~S.}\ \bibnamefont {Agarwal}},\ and\ \bibinfo {author} {\bibfnamefont {H.}~\bibnamefont {Walther}},\ }\bibfield  {title} {\bibinfo {title} {Extracting work from a single heat bath via vanishing quantum coherence},\ }\href {https://doi.org/10.1126/science.1078955} {\bibfield  {journal} {\bibinfo  {journal} {Science}\ }\textbf {\bibinfo {volume} {299}},\ \bibinfo {pages} {862} (\bibinfo {year} {2003})}\BibitemShut {NoStop}%
\bibitem [{\citenamefont {Agarwal}\ and\ \citenamefont {Puri}(1990)}]{agarwal1990cooperative}%
  \BibitemOpen
  \bibfield  {author} {\bibinfo {author} {\bibfnamefont {G.}~\bibnamefont {Agarwal}}\ and\ \bibinfo {author} {\bibfnamefont {R.}~\bibnamefont {Puri}},\ }\bibfield  {title} {\bibinfo {title} {Cooperative behavior of atoms irradiated by broadband squeezed light},\ }\href {https://doi.org/10.1103/PhysRevA.41.3782} {\bibfield  {journal} {\bibinfo  {journal} {Physical Review A}\ }\textbf {\bibinfo {volume} {41}},\ \bibinfo {pages} {3782} (\bibinfo {year} {1990})}\BibitemShut {NoStop}%
\bibitem [{\citenamefont {Dicke}(1954)}]{dicke1954coherence}%
  \BibitemOpen
  \bibfield  {author} {\bibinfo {author} {\bibfnamefont {R.~H.}\ \bibnamefont {Dicke}},\ }\bibfield  {title} {\bibinfo {title} {Coherence in spontaneous radiation processes},\ }\href {https://doi.org/10.1103/PhysRev.93.99} {\bibfield  {journal} {\bibinfo  {journal} {Physical Review}\ }\textbf {\bibinfo {volume} {93}},\ \bibinfo {pages} {99} (\bibinfo {year} {1954})}\BibitemShut {NoStop}%
\bibitem [{\citenamefont {Ferraro}\ \emph {et~al.}(2018)\citenamefont {Ferraro}, \citenamefont {Campisi}, \citenamefont {Andolina}, \citenamefont {Pellegrini},\ and\ \citenamefont {Polini}}]{ferraro2018high}%
  \BibitemOpen
  \bibfield  {author} {\bibinfo {author} {\bibfnamefont {D.}~\bibnamefont {Ferraro}}, \bibinfo {author} {\bibfnamefont {M.}~\bibnamefont {Campisi}}, \bibinfo {author} {\bibfnamefont {G.~M.}\ \bibnamefont {Andolina}}, \bibinfo {author} {\bibfnamefont {V.}~\bibnamefont {Pellegrini}},\ and\ \bibinfo {author} {\bibfnamefont {M.}~\bibnamefont {Polini}},\ }\bibfield  {title} {\bibinfo {title} {High-power collective charging of a solid-state quantum battery},\ }\href {https://doi.org/10.1103/PhysRevLett.120.117702} {\bibfield  {journal} {\bibinfo  {journal} {Physical Review Letters}\ }\textbf {\bibinfo {volume} {120}},\ \bibinfo {pages} {117702} (\bibinfo {year} {2018})}\BibitemShut {NoStop}%
\bibitem [{\citenamefont {Campaioli}\ \emph {et~al.}(2017)\citenamefont {Campaioli}, \citenamefont {Pollock}, \citenamefont {Binder}, \citenamefont {C{\'e}leri}, \citenamefont {Goold}, \citenamefont {Vinjanampathy},\ and\ \citenamefont {Modi}}]{campaioli2017enhancing}%
  \BibitemOpen
  \bibfield  {author} {\bibinfo {author} {\bibfnamefont {F.}~\bibnamefont {Campaioli}}, \bibinfo {author} {\bibfnamefont {F.~A.}\ \bibnamefont {Pollock}}, \bibinfo {author} {\bibfnamefont {F.~C.}\ \bibnamefont {Binder}}, \bibinfo {author} {\bibfnamefont {L.}~\bibnamefont {C{\'e}leri}}, \bibinfo {author} {\bibfnamefont {J.}~\bibnamefont {Goold}}, \bibinfo {author} {\bibfnamefont {S.}~\bibnamefont {Vinjanampathy}},\ and\ \bibinfo {author} {\bibfnamefont {K.}~\bibnamefont {Modi}},\ }\bibfield  {title} {\bibinfo {title} {Enhancing the charging power of quantum batteries},\ }\href {https://doi.org/10.1103/PhysRevLett.118.150601} {\bibfield  {journal} {\bibinfo  {journal} {Physical Review Letters}\ }\textbf {\bibinfo {volume} {118}},\ \bibinfo {pages} {150601} (\bibinfo {year} {2017})}\BibitemShut {NoStop}%
\bibitem [{\citenamefont {Alicki}\ and\ \citenamefont {Fannes}(2013)}]{alicki2013entanglement}%
  \BibitemOpen
  \bibfield  {author} {\bibinfo {author} {\bibfnamefont {R.}~\bibnamefont {Alicki}}\ and\ \bibinfo {author} {\bibfnamefont {M.}~\bibnamefont {Fannes}},\ }\bibfield  {title} {\bibinfo {title} {Entanglement boost for extractable work from ensembles of quantum batteries},\ }\href {https://doi.org/10.1103/PhysRevE.87.042123} {\bibfield  {journal} {\bibinfo  {journal} {Physical Review E—Statistical, Nonlinear, and Soft Matter Physics}\ }\textbf {\bibinfo {volume} {87}},\ \bibinfo {pages} {042123} (\bibinfo {year} {2013})}\BibitemShut {NoStop}%
\bibitem [{\citenamefont {Campaioli}\ \emph {et~al.}(2024)\citenamefont {Campaioli}, \citenamefont {Gherardini}, \citenamefont {Quach}, \citenamefont {Polini},\ and\ \citenamefont {Andolina}}]{campaioli2024colloquium}%
  \BibitemOpen
  \bibfield  {author} {\bibinfo {author} {\bibfnamefont {F.}~\bibnamefont {Campaioli}}, \bibinfo {author} {\bibfnamefont {S.}~\bibnamefont {Gherardini}}, \bibinfo {author} {\bibfnamefont {J.~Q.}\ \bibnamefont {Quach}}, \bibinfo {author} {\bibfnamefont {M.}~\bibnamefont {Polini}},\ and\ \bibinfo {author} {\bibfnamefont {G.~M.}\ \bibnamefont {Andolina}},\ }\bibfield  {title} {\bibinfo {title} {Colloquium: quantum batteries},\ }\href {https://doi.org/10.1103/RevModPhys.96.031001} {\bibfield  {journal} {\bibinfo  {journal} {Reviews of Modern Physics}\ }\textbf {\bibinfo {volume} {96}},\ \bibinfo {pages} {031001} (\bibinfo {year} {2024})}\BibitemShut {NoStop}%
\bibitem [{\citenamefont {Ferraro}\ \emph {et~al.}(2026)\citenamefont {Ferraro}, \citenamefont {Cavaliere}, \citenamefont {Genoni}, \citenamefont {Benenti},\ and\ \citenamefont {Sassetti}}]{ferraro2026opportunities}%
  \BibitemOpen
  \bibfield  {author} {\bibinfo {author} {\bibfnamefont {D.}~\bibnamefont {Ferraro}}, \bibinfo {author} {\bibfnamefont {F.}~\bibnamefont {Cavaliere}}, \bibinfo {author} {\bibfnamefont {M.~G.}\ \bibnamefont {Genoni}}, \bibinfo {author} {\bibfnamefont {G.}~\bibnamefont {Benenti}},\ and\ \bibinfo {author} {\bibfnamefont {M.}~\bibnamefont {Sassetti}},\ }\bibfield  {title} {\bibinfo {title} {Opportunities and challenges of quantum batteries},\ }\href {https://doi.org/10.1038/s42254-025-00906-5} {\bibfield  {journal} {\bibinfo  {journal} {Nature Reviews Physics}\ ,\ \bibinfo {pages} {1}} (\bibinfo {year} {2026})}\BibitemShut {NoStop}%
\bibitem [{\citenamefont {Andolina}\ \emph {et~al.}(2019)\citenamefont {Andolina}, \citenamefont {Keck}, \citenamefont {Mari}, \citenamefont {Campisi}, \citenamefont {Giovannetti},\ and\ \citenamefont {Polini}}]{andolina2019extractable}%
  \BibitemOpen
  \bibfield  {author} {\bibinfo {author} {\bibfnamefont {G.~M.}\ \bibnamefont {Andolina}}, \bibinfo {author} {\bibfnamefont {M.}~\bibnamefont {Keck}}, \bibinfo {author} {\bibfnamefont {A.}~\bibnamefont {Mari}}, \bibinfo {author} {\bibfnamefont {M.}~\bibnamefont {Campisi}}, \bibinfo {author} {\bibfnamefont {V.}~\bibnamefont {Giovannetti}},\ and\ \bibinfo {author} {\bibfnamefont {M.}~\bibnamefont {Polini}},\ }\bibfield  {title} {\bibinfo {title} {Extractable work, the role of correlations, and asymptotic freedom in quantum batteries},\ }\href {https://doi.org/10.1103/PhysRevLett.122.047702} {\bibfield  {journal} {\bibinfo  {journal} {Physical review letters}\ }\textbf {\bibinfo {volume} {122}},\ \bibinfo {pages} {047702} (\bibinfo {year} {2019})}\BibitemShut {NoStop}%
\bibitem [{\citenamefont {Kamin}\ \emph {et~al.}(2020)\citenamefont {Kamin}, \citenamefont {Tabesh}, \citenamefont {Salimi},\ and\ \citenamefont {Santos}}]{kamin2020entanglement}%
  \BibitemOpen
  \bibfield  {author} {\bibinfo {author} {\bibfnamefont {F.}~\bibnamefont {Kamin}}, \bibinfo {author} {\bibfnamefont {F.}~\bibnamefont {Tabesh}}, \bibinfo {author} {\bibfnamefont {S.}~\bibnamefont {Salimi}},\ and\ \bibinfo {author} {\bibfnamefont {A.~C.}\ \bibnamefont {Santos}},\ }\bibfield  {title} {\bibinfo {title} {Entanglement, coherence, and charging process of quantum batteries},\ }\href {https://doi.org/10.1103/PhysRevE.102.052109} {\bibfield  {journal} {\bibinfo  {journal} {Physical Review E}\ }\textbf {\bibinfo {volume} {102}},\ \bibinfo {pages} {052109} (\bibinfo {year} {2020})}\BibitemShut {NoStop}%
\bibitem [{\citenamefont {Tabesh}\ \emph {et~al.}(2020)\citenamefont {Tabesh}, \citenamefont {Kamin},\ and\ \citenamefont {Salimi}}]{tabesh2020environment}%
  \BibitemOpen
  \bibfield  {author} {\bibinfo {author} {\bibfnamefont {F.}~\bibnamefont {Tabesh}}, \bibinfo {author} {\bibfnamefont {F.}~\bibnamefont {Kamin}},\ and\ \bibinfo {author} {\bibfnamefont {S.}~\bibnamefont {Salimi}},\ }\bibfield  {title} {\bibinfo {title} {Environment-mediated charging process of quantum batteries},\ }\href {https://doi.org/10.1103/PhysRevA.102.052223} {\bibfield  {journal} {\bibinfo  {journal} {Physical Review A}\ }\textbf {\bibinfo {volume} {102}},\ \bibinfo {pages} {052223} (\bibinfo {year} {2020})}\BibitemShut {NoStop}%
\bibitem [{\citenamefont {Ferreri}\ \emph {et~al.}(2025)\citenamefont {Ferreri}, \citenamefont {Wang}, \citenamefont {Nori}, \citenamefont {Wilhelm},\ and\ \citenamefont {Bruschi}}]{ferreri2025quantum}%
  \BibitemOpen
  \bibfield  {author} {\bibinfo {author} {\bibfnamefont {A.}~\bibnamefont {Ferreri}}, \bibinfo {author} {\bibfnamefont {H.}~\bibnamefont {Wang}}, \bibinfo {author} {\bibfnamefont {F.}~\bibnamefont {Nori}}, \bibinfo {author} {\bibfnamefont {F.~K.}\ \bibnamefont {Wilhelm}},\ and\ \bibinfo {author} {\bibfnamefont {D.~E.}\ \bibnamefont {Bruschi}},\ }\bibfield  {title} {\bibinfo {title} {Quantum heat engine based on quantum interferometry: The {SU} (1, 1) {O}tto cycle},\ }\href {https://doi.org/10.1103/PhysRevResearch.7.013284} {\bibfield  {journal} {\bibinfo  {journal} {Physical Review Research}\ }\textbf {\bibinfo {volume} {7}},\ \bibinfo {pages} {013284} (\bibinfo {year} {2025})}\BibitemShut {NoStop}%
\bibitem [{\citenamefont {Centrone}\ \emph {et~al.}(2023)\citenamefont {Centrone}, \citenamefont {Mancino},\ and\ \citenamefont {Paternostro}}]{centrone2023charging}%
  \BibitemOpen
  \bibfield  {author} {\bibinfo {author} {\bibfnamefont {F.}~\bibnamefont {Centrone}}, \bibinfo {author} {\bibfnamefont {L.}~\bibnamefont {Mancino}},\ and\ \bibinfo {author} {\bibfnamefont {M.}~\bibnamefont {Paternostro}},\ }\bibfield  {title} {\bibinfo {title} {Charging batteries with quantum squeezing},\ }\href {https://doi.org/10.1103/PhysRevA.108.052213} {\bibfield  {journal} {\bibinfo  {journal} {Physical Review A}\ }\textbf {\bibinfo {volume} {108}},\ \bibinfo {pages} {052213} (\bibinfo {year} {2023})}\BibitemShut {NoStop}%
\bibitem [{\citenamefont {Li}\ \emph {et~al.}(2025)\citenamefont {Li}, \citenamefont {Ma}, \citenamefont {Hao},\ and\ \citenamefont {Yu}}]{li2025enhancing}%
  \BibitemOpen
  \bibfield  {author} {\bibinfo {author} {\bibfnamefont {H.}~\bibnamefont {Li}}, \bibinfo {author} {\bibfnamefont {H.}~\bibnamefont {Ma}}, \bibinfo {author} {\bibfnamefont {Y.}~\bibnamefont {Hao}},\ and\ \bibinfo {author} {\bibfnamefont {W.}~\bibnamefont {Yu}},\ }\bibfield  {title} {\bibinfo {title} {Enhancing ergotropy of a quantum battery with coherent chargers: The catalystlike role of indefinite causal order},\ }\href {https://doi.org/10.1103/s6dl-zgkx} {\bibfield  {journal} {\bibinfo  {journal} {Physical Review A}\ }\textbf {\bibinfo {volume} {112}},\ \bibinfo {pages} {042229} (\bibinfo {year} {2025})}\BibitemShut {NoStop}%
\bibitem [{\citenamefont {Lai}\ \emph {et~al.}(2024)\citenamefont {Lai}, \citenamefont {Lin}, \citenamefont {Huang}, \citenamefont {Jan},\ and\ \citenamefont {Chen}}]{lai2024quick}%
  \BibitemOpen
  \bibfield  {author} {\bibinfo {author} {\bibfnamefont {P.-R.}\ \bibnamefont {Lai}}, \bibinfo {author} {\bibfnamefont {J.-D.}\ \bibnamefont {Lin}}, \bibinfo {author} {\bibfnamefont {Y.-T.}\ \bibnamefont {Huang}}, \bibinfo {author} {\bibfnamefont {H.-C.}\ \bibnamefont {Jan}},\ and\ \bibinfo {author} {\bibfnamefont {Y.-N.}\ \bibnamefont {Chen}},\ }\bibfield  {title} {\bibinfo {title} {Quick charging of a quantum battery with superposed trajectories},\ }\href {https://doi.org/10.1103/PhysRevResearch.6.023136} {\bibfield  {journal} {\bibinfo  {journal} {Physical Review Research}\ }\textbf {\bibinfo {volume} {6}},\ \bibinfo {pages} {023136} (\bibinfo {year} {2024})}\BibitemShut {NoStop}%
\bibitem [{\citenamefont {Allahverdyan}\ \emph {et~al.}(2004)\citenamefont {Allahverdyan}, \citenamefont {Balian},\ and\ \citenamefont {Nieuwenhuizen}}]{allahverdyan2004maximal}%
  \BibitemOpen
  \bibfield  {author} {\bibinfo {author} {\bibfnamefont {A.~E.}\ \bibnamefont {Allahverdyan}}, \bibinfo {author} {\bibfnamefont {R.}~\bibnamefont {Balian}},\ and\ \bibinfo {author} {\bibfnamefont {T.~M.}\ \bibnamefont {Nieuwenhuizen}},\ }\bibfield  {title} {\bibinfo {title} {Maximal work extraction from finite quantum systems},\ }\href {https://doi.org/10.1209/epl/i2004-10101-2} {\bibfield  {journal} {\bibinfo  {journal} {Europhysics Letters}\ }\textbf {\bibinfo {volume} {67}},\ \bibinfo {pages} {565} (\bibinfo {year} {2004})}\BibitemShut {NoStop}%
\bibitem [{\citenamefont {Francica}\ \emph {et~al.}(2020)\citenamefont {Francica}, \citenamefont {Binder}, \citenamefont {Guarnieri}, \citenamefont {Mitchison}, \citenamefont {Goold},\ and\ \citenamefont {Plastina}}]{francica2020quantum}%
  \BibitemOpen
  \bibfield  {author} {\bibinfo {author} {\bibfnamefont {G.}~\bibnamefont {Francica}}, \bibinfo {author} {\bibfnamefont {F.~C.}\ \bibnamefont {Binder}}, \bibinfo {author} {\bibfnamefont {G.}~\bibnamefont {Guarnieri}}, \bibinfo {author} {\bibfnamefont {M.~T.}\ \bibnamefont {Mitchison}}, \bibinfo {author} {\bibfnamefont {J.}~\bibnamefont {Goold}},\ and\ \bibinfo {author} {\bibfnamefont {F.}~\bibnamefont {Plastina}},\ }\bibfield  {title} {\bibinfo {title} {Quantum coherence and ergotropy},\ }\href {https://doi.org/10.1103/PhysRevLett.125.180603} {\bibfield  {journal} {\bibinfo  {journal} {Physical Review Letters}\ }\textbf {\bibinfo {volume} {125}},\ \bibinfo {pages} {180603} (\bibinfo {year} {2020})}\BibitemShut {NoStop}%
\bibitem [{\citenamefont {Liu}\ \emph {et~al.}(2019)\citenamefont {Liu}, \citenamefont {Segal},\ and\ \citenamefont {Hanna}}]{liu2019loss}%
  \BibitemOpen
  \bibfield  {author} {\bibinfo {author} {\bibfnamefont {J.}~\bibnamefont {Liu}}, \bibinfo {author} {\bibfnamefont {D.}~\bibnamefont {Segal}},\ and\ \bibinfo {author} {\bibfnamefont {G.}~\bibnamefont {Hanna}},\ }\bibfield  {title} {\bibinfo {title} {Loss-free excitonic quantum battery},\ }\href {https://doi.org/10.1021/acs.jpcc.9b06373} {\bibfield  {journal} {\bibinfo  {journal} {The Journal of Physical Chemistry C}\ }\textbf {\bibinfo {volume} {123}},\ \bibinfo {pages} {18303} (\bibinfo {year} {2019})}\BibitemShut {NoStop}%
\bibitem [{\citenamefont {Gherardini}\ \emph {et~al.}(2020)\citenamefont {Gherardini}, \citenamefont {Campaioli}, \citenamefont {Caruso},\ and\ \citenamefont {Binder}}]{gherardini2020stabilizing}%
  \BibitemOpen
  \bibfield  {author} {\bibinfo {author} {\bibfnamefont {S.}~\bibnamefont {Gherardini}}, \bibinfo {author} {\bibfnamefont {F.}~\bibnamefont {Campaioli}}, \bibinfo {author} {\bibfnamefont {F.}~\bibnamefont {Caruso}},\ and\ \bibinfo {author} {\bibfnamefont {F.~C.}\ \bibnamefont {Binder}},\ }\bibfield  {title} {\bibinfo {title} {Stabilizing open quantum batteries by sequential measurements},\ }\href {https://doi.org/10.1103/PhysRevResearch.2.013095} {\bibfield  {journal} {\bibinfo  {journal} {Physical Review Research}\ }\textbf {\bibinfo {volume} {2}},\ \bibinfo {pages} {013095} (\bibinfo {year} {2020})}\BibitemShut {NoStop}%
\bibitem [{\citenamefont {Mitchison}\ \emph {et~al.}(2021)\citenamefont {Mitchison}, \citenamefont {Goold},\ and\ \citenamefont {Prior}}]{mitchison2021charging}%
  \BibitemOpen
  \bibfield  {author} {\bibinfo {author} {\bibfnamefont {M.~T.}\ \bibnamefont {Mitchison}}, \bibinfo {author} {\bibfnamefont {J.}~\bibnamefont {Goold}},\ and\ \bibinfo {author} {\bibfnamefont {J.}~\bibnamefont {Prior}},\ }\bibfield  {title} {\bibinfo {title} {Charging a quantum battery with linear feedback control},\ }\href {https://doi.org/10.1103/PhysRevResearch.2.013095} {\bibfield  {journal} {\bibinfo  {journal} {Quantum}\ }\textbf {\bibinfo {volume} {5}},\ \bibinfo {pages} {500} (\bibinfo {year} {2021})}\BibitemShut {NoStop}%
\bibitem [{\citenamefont {Pirmoradian}\ and\ \citenamefont {M{\o}lmer}(2019)}]{pirmoradian2019aging}%
  \BibitemOpen
  \bibfield  {author} {\bibinfo {author} {\bibfnamefont {F.}~\bibnamefont {Pirmoradian}}\ and\ \bibinfo {author} {\bibfnamefont {K.}~\bibnamefont {M{\o}lmer}},\ }\bibfield  {title} {\bibinfo {title} {Aging of a quantum battery},\ }\href {https://doi.org/10.1103/PhysRevA.100.043833} {\bibfield  {journal} {\bibinfo  {journal} {Physical Review A}\ }\textbf {\bibinfo {volume} {100}},\ \bibinfo {pages} {043833} (\bibinfo {year} {2019})}\BibitemShut {NoStop}%
\bibitem [{\citenamefont {Quach}\ and\ \citenamefont {Munro}(2020)}]{quach2020using}%
  \BibitemOpen
  \bibfield  {author} {\bibinfo {author} {\bibfnamefont {J.~Q.}\ \bibnamefont {Quach}}\ and\ \bibinfo {author} {\bibfnamefont {W.~J.}\ \bibnamefont {Munro}},\ }\bibfield  {title} {\bibinfo {title} {Using {Dark} {States} to {Charge} and {Stabilize} {Open} {Quantum} {Batteries}},\ }\href {https://doi.org/10.1103/PhysRevApplied.14.024092} {\bibfield  {journal} {\bibinfo  {journal} {Phys. Rev. Appl.}\ }\textbf {\bibinfo {volume} {14}},\ \bibinfo {pages} {024092} (\bibinfo {year} {2020})}\BibitemShut {NoStop}%
\bibitem [{\citenamefont {Gyhm}\ \emph {et~al.}(2022)\citenamefont {Gyhm}, \citenamefont {{\v{S}}afr{\'a}nek},\ and\ \citenamefont {Rosa}}]{gyhm2022quantum}%
  \BibitemOpen
  \bibfield  {author} {\bibinfo {author} {\bibfnamefont {J.-Y.}\ \bibnamefont {Gyhm}}, \bibinfo {author} {\bibfnamefont {D.}~\bibnamefont {{\v{S}}afr{\'a}nek}},\ and\ \bibinfo {author} {\bibfnamefont {D.}~\bibnamefont {Rosa}},\ }\bibfield  {title} {\bibinfo {title} {Quantum charging advantage cannot be extensive without global operations},\ }\href {https://doi.org/10.1103/PhysRevLett.128.140501} {\bibfield  {journal} {\bibinfo  {journal} {Physical Review Letters}\ }\textbf {\bibinfo {volume} {128}},\ \bibinfo {pages} {140501} (\bibinfo {year} {2022})}\BibitemShut {NoStop}%
\bibitem [{\citenamefont {Binder}\ \emph {et~al.}(2015)\citenamefont {Binder}, \citenamefont {Vinjanampathy}, \citenamefont {Modi},\ and\ \citenamefont {Goold}}]{binder2015quantacell}%
  \BibitemOpen
  \bibfield  {author} {\bibinfo {author} {\bibfnamefont {F.~C.}\ \bibnamefont {Binder}}, \bibinfo {author} {\bibfnamefont {S.}~\bibnamefont {Vinjanampathy}}, \bibinfo {author} {\bibfnamefont {K.}~\bibnamefont {Modi}},\ and\ \bibinfo {author} {\bibfnamefont {J.}~\bibnamefont {Goold}},\ }\bibfield  {title} {\bibinfo {title} {Quantacell: powerful charging of quantum batteries},\ }\href {https://doi.org/10.1088/1367-2630/17/7/075015} {\bibfield  {journal} {\bibinfo  {journal} {New Journal of Physics}\ }\textbf {\bibinfo {volume} {17}},\ \bibinfo {pages} {075015} (\bibinfo {year} {2015})}\BibitemShut {NoStop}%
\bibitem [{\citenamefont {Rodr{\'\i}guez}\ \emph {et~al.}(2024)\citenamefont {Rodr{\'\i}guez}, \citenamefont {Ahmadi}, \citenamefont {Su{\'a}rez}, \citenamefont {Mazurek}, \citenamefont {Barzanjeh},\ and\ \citenamefont {Horodecki}}]{rodriguez2024optimal}%
  \BibitemOpen
  \bibfield  {author} {\bibinfo {author} {\bibfnamefont {R.~R.}\ \bibnamefont {Rodr{\'\i}guez}}, \bibinfo {author} {\bibfnamefont {B.}~\bibnamefont {Ahmadi}}, \bibinfo {author} {\bibfnamefont {G.}~\bibnamefont {Su{\'a}rez}}, \bibinfo {author} {\bibfnamefont {P.}~\bibnamefont {Mazurek}}, \bibinfo {author} {\bibfnamefont {S.}~\bibnamefont {Barzanjeh}},\ and\ \bibinfo {author} {\bibfnamefont {P.}~\bibnamefont {Horodecki}},\ }\bibfield  {title} {\bibinfo {title} {Optimal quantum control of charging quantum batteries},\ }\href {https://doi.org/10.1088/1367-2630/ad3843} {\bibfield  {journal} {\bibinfo  {journal} {New Journal of Physics}\ }\textbf {\bibinfo {volume} {26}},\ \bibinfo {pages} {043004} (\bibinfo {year} {2024})}\BibitemShut {NoStop}%
\bibitem [{\citenamefont {Auff{\`e}ves}(2022)}]{auffeves2022quantum}%
  \BibitemOpen
  \bibfield  {author} {\bibinfo {author} {\bibfnamefont {A.}~\bibnamefont {Auff{\`e}ves}},\ }\bibfield  {title} {\bibinfo {title} {Quantum technologies need a quantum energy initiative},\ }\href {https://doi.org/10.1103/PRXQuantum.3.020101} {\bibfield  {journal} {\bibinfo  {journal} {PRX Quantum}\ }\textbf {\bibinfo {volume} {3}},\ \bibinfo {pages} {020101} (\bibinfo {year} {2022})}\BibitemShut {NoStop}%
\bibitem [{\citenamefont {Fellous-Asiani}\ \emph {et~al.}(2023)\citenamefont {Fellous-Asiani}, \citenamefont {Chai}, \citenamefont {Thonnart}, \citenamefont {Ng}, \citenamefont {Whitney},\ and\ \citenamefont {Auff{\`e}ves}}]{fellous2023optimizing}%
  \BibitemOpen
  \bibfield  {author} {\bibinfo {author} {\bibfnamefont {M.}~\bibnamefont {Fellous-Asiani}}, \bibinfo {author} {\bibfnamefont {J.~H.}\ \bibnamefont {Chai}}, \bibinfo {author} {\bibfnamefont {Y.}~\bibnamefont {Thonnart}}, \bibinfo {author} {\bibfnamefont {H.~K.}\ \bibnamefont {Ng}}, \bibinfo {author} {\bibfnamefont {R.~S.}\ \bibnamefont {Whitney}},\ and\ \bibinfo {author} {\bibfnamefont {A.}~\bibnamefont {Auff{\`e}ves}},\ }\bibfield  {title} {\bibinfo {title} {Optimizing resource efficiencies for scalable full-stack quantum computers},\ }\href {https://doi.org/10.1103/PRXQuantum.4.040319} {\bibfield  {journal} {\bibinfo  {journal} {PRX Quantum}\ }\textbf {\bibinfo {volume} {4}},\ \bibinfo {pages} {040319} (\bibinfo {year} {2023})}\BibitemShut {NoStop}%
\bibitem [{\citenamefont {Mayo}\ and\ \citenamefont {Roncaglia}(2022)}]{mayo2022collective}%
  \BibitemOpen
  \bibfield  {author} {\bibinfo {author} {\bibfnamefont {F.}~\bibnamefont {Mayo}}\ and\ \bibinfo {author} {\bibfnamefont {A.~J.}\ \bibnamefont {Roncaglia}},\ }\bibfield  {title} {\bibinfo {title} {Collective effects and quantum coherence in dissipative charging of quantum batteries},\ }\href {https://doi.org/10.1103/PhysRevA.105.062203} {\bibfield  {journal} {\bibinfo  {journal} {Physical Review A}\ }\textbf {\bibinfo {volume} {105}},\ \bibinfo {pages} {062203} (\bibinfo {year} {2022})}\BibitemShut {NoStop}%
\bibitem [{\citenamefont {Leroux}\ \emph {et~al.}(2010)\citenamefont {Leroux}, \citenamefont {Schleier-Smith},\ and\ \citenamefont {Vuleti{\'c}}}]{leroux2010implementation}%
  \BibitemOpen
  \bibfield  {author} {\bibinfo {author} {\bibfnamefont {I.~D.}\ \bibnamefont {Leroux}}, \bibinfo {author} {\bibfnamefont {M.~H.}\ \bibnamefont {Schleier-Smith}},\ and\ \bibinfo {author} {\bibfnamefont {V.}~\bibnamefont {Vuleti{\'c}}},\ }\bibfield  {title} {\bibinfo {title} {Implementation of cavity squeezing of a collective atomic spin},\ }\href {https://doi.org/10.1103/PhysRevLett.104.073602} {\bibfield  {journal} {\bibinfo  {journal} {Physical Review Letters}\ }\textbf {\bibinfo {volume} {104}},\ \bibinfo {pages} {073602} (\bibinfo {year} {2010})}\BibitemShut {NoStop}%
\bibitem [{\citenamefont {Byrnes}\ \emph {et~al.}(2015)\citenamefont {Byrnes}, \citenamefont {Rosseau}, \citenamefont {Khosla}, \citenamefont {Pyrkov}, \citenamefont {Thomasen}, \citenamefont {Mukai}, \citenamefont {Koyama}, \citenamefont {Abdelrahman},\ and\ \citenamefont {Ilo-Okeke}}]{byrnes2015macroscopic}%
  \BibitemOpen
  \bibfield  {author} {\bibinfo {author} {\bibfnamefont {T.}~\bibnamefont {Byrnes}}, \bibinfo {author} {\bibfnamefont {D.}~\bibnamefont {Rosseau}}, \bibinfo {author} {\bibfnamefont {M.}~\bibnamefont {Khosla}}, \bibinfo {author} {\bibfnamefont {A.}~\bibnamefont {Pyrkov}}, \bibinfo {author} {\bibfnamefont {A.}~\bibnamefont {Thomasen}}, \bibinfo {author} {\bibfnamefont {T.}~\bibnamefont {Mukai}}, \bibinfo {author} {\bibfnamefont {S.}~\bibnamefont {Koyama}}, \bibinfo {author} {\bibfnamefont {A.}~\bibnamefont {Abdelrahman}},\ and\ \bibinfo {author} {\bibfnamefont {E.}~\bibnamefont {Ilo-Okeke}},\ }\bibfield  {title} {\bibinfo {title} {Macroscopic quantum information processing using spin coherent states},\ }\href {https://doi.org/https://doi.org/10.1016/j.optcom.2014.08.017} {\bibfield  {journal} {\bibinfo  {journal} {Optics Communications}\ }\textbf {\bibinfo {volume} {337}},\ \bibinfo {pages} {102} (\bibinfo {year} {2015})}\BibitemShut {NoStop}%
\bibitem [{\citenamefont {Auccaise~Estrada}\ \emph {et~al.}(2013)\citenamefont {Auccaise~Estrada}, \citenamefont {de~Azevedo}, \citenamefont {Duzzioni}, \citenamefont {Bonagamba},\ and\ \citenamefont {Youssef~Moussa}}]{auccaise2013spin}%
  \BibitemOpen
  \bibfield  {author} {\bibinfo {author} {\bibfnamefont {R.}~\bibnamefont {Auccaise~Estrada}}, \bibinfo {author} {\bibfnamefont {E.~R.}\ \bibnamefont {de~Azevedo}}, \bibinfo {author} {\bibfnamefont {E.~I.}\ \bibnamefont {Duzzioni}}, \bibinfo {author} {\bibfnamefont {T.~J.}\ \bibnamefont {Bonagamba}},\ and\ \bibinfo {author} {\bibfnamefont {M.~H.}\ \bibnamefont {Youssef~Moussa}},\ }\bibfield  {title} {\bibinfo {title} {Spin coherent states in {NMR} quadrupolar system: experimental and theoretical applications},\ }\href {https://doi.org/10.1140/epjd/e2013-30689-1} {\bibfield  {journal} {\bibinfo  {journal} {The European Physical Journal D}\ }\textbf {\bibinfo {volume} {67}},\ \bibinfo {pages} {127} (\bibinfo {year} {2013})}\BibitemShut {NoStop}%
\bibitem [{\citenamefont {Joshi}\ and\ \citenamefont {Mahesh}(2022)}]{joshi2022experimental}%
  \BibitemOpen
  \bibfield  {author} {\bibinfo {author} {\bibfnamefont {J.}~\bibnamefont {Joshi}}\ and\ \bibinfo {author} {\bibfnamefont {T.}~\bibnamefont {Mahesh}},\ }\bibfield  {title} {\bibinfo {title} {Experimental investigation of a quantum battery using star-topology nmr spin systems},\ }\href {https://doi.org/10.1103/PhysRevA.106.042601} {\bibfield  {journal} {\bibinfo  {journal} {Physical Review A}\ }\textbf {\bibinfo {volume} {106}},\ \bibinfo {pages} {042601} (\bibinfo {year} {2022})}\BibitemShut {NoStop}%
\bibitem [{\citenamefont {Angerer}\ \emph {et~al.}(2018)\citenamefont {Angerer}, \citenamefont {Streltsov}, \citenamefont {Astner}, \citenamefont {Putz}, \citenamefont {Sumiya}, \citenamefont {Onoda}, \citenamefont {Isoya}, \citenamefont {Munro}, \citenamefont {Nemoto}, \citenamefont {Schmiedmayer} \emph {et~al.}}]{angerer2018superradiant}%
  \BibitemOpen
  \bibfield  {author} {\bibinfo {author} {\bibfnamefont {A.}~\bibnamefont {Angerer}}, \bibinfo {author} {\bibfnamefont {K.}~\bibnamefont {Streltsov}}, \bibinfo {author} {\bibfnamefont {T.}~\bibnamefont {Astner}}, \bibinfo {author} {\bibfnamefont {S.}~\bibnamefont {Putz}}, \bibinfo {author} {\bibfnamefont {H.}~\bibnamefont {Sumiya}}, \bibinfo {author} {\bibfnamefont {S.}~\bibnamefont {Onoda}}, \bibinfo {author} {\bibfnamefont {J.}~\bibnamefont {Isoya}}, \bibinfo {author} {\bibfnamefont {W.~J.}\ \bibnamefont {Munro}}, \bibinfo {author} {\bibfnamefont {K.}~\bibnamefont {Nemoto}}, \bibinfo {author} {\bibfnamefont {J.}~\bibnamefont {Schmiedmayer}}, \emph {et~al.},\ }\bibfield  {title} {\bibinfo {title} {Superradiant emission from colour centres in diamond},\ }\href {https://doi.org/10.1038/s41567-018-0269-7} {\bibfield  {journal} {\bibinfo  {journal} {Nature Physics}\ }\textbf {\bibinfo {volume} {14}},\ \bibinfo {pages} {1168} (\bibinfo {year} {2018})}\BibitemShut {NoStop}%
\bibitem [{\citenamefont {Murch}\ \emph {et~al.}(2013)\citenamefont {Murch}, \citenamefont {Weber}, \citenamefont {Beck}, \citenamefont {Ginossar},\ and\ \citenamefont {Siddiqi}}]{murch2013reduction}%
  \BibitemOpen
  \bibfield  {author} {\bibinfo {author} {\bibfnamefont {K.}~\bibnamefont {Murch}}, \bibinfo {author} {\bibfnamefont {S.}~\bibnamefont {Weber}}, \bibinfo {author} {\bibfnamefont {K.}~\bibnamefont {Beck}}, \bibinfo {author} {\bibfnamefont {E.}~\bibnamefont {Ginossar}},\ and\ \bibinfo {author} {\bibfnamefont {I.}~\bibnamefont {Siddiqi}},\ }\bibfield  {title} {\bibinfo {title} {Reduction of the radiative decay of atomic coherence in squeezed vacuum},\ }\href {https://doi.org/10.1038/nature12264} {\bibfield  {journal} {\bibinfo  {journal} {Nature}\ }\textbf {\bibinfo {volume} {499}},\ \bibinfo {pages} {62} (\bibinfo {year} {2013})}\BibitemShut {NoStop}%
\bibitem [{\citenamefont {Mallet}\ \emph {et~al.}(2011)\citenamefont {Mallet}, \citenamefont {Castellanos-Beltran}, \citenamefont {Ku}, \citenamefont {Glancy}, \citenamefont {Knill}, \citenamefont {Irwin}, \citenamefont {Hilton}, \citenamefont {Vale},\ and\ \citenamefont {Lehnert}}]{mallet2011quantum}%
  \BibitemOpen
  \bibfield  {author} {\bibinfo {author} {\bibfnamefont {F.}~\bibnamefont {Mallet}}, \bibinfo {author} {\bibfnamefont {M.}~\bibnamefont {Castellanos-Beltran}}, \bibinfo {author} {\bibfnamefont {H.}~\bibnamefont {Ku}}, \bibinfo {author} {\bibfnamefont {S.}~\bibnamefont {Glancy}}, \bibinfo {author} {\bibfnamefont {E.}~\bibnamefont {Knill}}, \bibinfo {author} {\bibfnamefont {K.}~\bibnamefont {Irwin}}, \bibinfo {author} {\bibfnamefont {G.}~\bibnamefont {Hilton}}, \bibinfo {author} {\bibfnamefont {L.}~\bibnamefont {Vale}},\ and\ \bibinfo {author} {\bibfnamefont {K.}~\bibnamefont {Lehnert}},\ }\bibfield  {title} {\bibinfo {title} {Quantum state tomography of an itinerant squeezed microwave field},\ }\href {https://doi.org/10.1103/PhysRevLett.106.220502} {\bibfield  {journal} {\bibinfo  {journal} {Physical Review Letters}\ }\textbf {\bibinfo {volume} {106}},\ \bibinfo {pages} {220502} (\bibinfo {year} {2011})}\BibitemShut {NoStop}%
\bibitem [{\citenamefont {Toyli}\ \emph {et~al.}(2016)\citenamefont {Toyli}, \citenamefont {Eddins}, \citenamefont {Boutin}, \citenamefont {Puri}, \citenamefont {Hover}, \citenamefont {Bolkhovsky}, \citenamefont {Oliver}, \citenamefont {Blais},\ and\ \citenamefont {Siddiqi}}]{toyli2016resonance}%
  \BibitemOpen
  \bibfield  {author} {\bibinfo {author} {\bibfnamefont {D.}~\bibnamefont {Toyli}}, \bibinfo {author} {\bibfnamefont {A.}~\bibnamefont {Eddins}}, \bibinfo {author} {\bibfnamefont {S.}~\bibnamefont {Boutin}}, \bibinfo {author} {\bibfnamefont {S.}~\bibnamefont {Puri}}, \bibinfo {author} {\bibfnamefont {D.}~\bibnamefont {Hover}}, \bibinfo {author} {\bibfnamefont {V.}~\bibnamefont {Bolkhovsky}}, \bibinfo {author} {\bibfnamefont {W.}~\bibnamefont {Oliver}}, \bibinfo {author} {\bibfnamefont {A.}~\bibnamefont {Blais}},\ and\ \bibinfo {author} {\bibfnamefont {I.}~\bibnamefont {Siddiqi}},\ }\bibfield  {title} {\bibinfo {title} {Resonance fluorescence from an artificial atom in squeezed vacuum},\ }\href {https://doi.org/10.1103/PhysRevX.6.031004} {\bibfield  {journal} {\bibinfo  {journal} {Physical Review X}\ }\textbf {\bibinfo {volume} {6}},\ \bibinfo {pages} {031004} (\bibinfo {year} {2016})}\BibitemShut {NoStop}%
\bibitem [{\citenamefont {Quach}\ \emph {et~al.}(2022)\citenamefont {Quach}, \citenamefont {McGhee}, \citenamefont {Ganzer}, \citenamefont {Rouse}, \citenamefont {Lovett}, \citenamefont {Gauger}, \citenamefont {Keeling}, \citenamefont {Cerullo}, \citenamefont {Lidzey},\ and\ \citenamefont {Virgili}}]{quach2022superabsorption}%
  \BibitemOpen
  \bibfield  {author} {\bibinfo {author} {\bibfnamefont {J.~Q.}\ \bibnamefont {Quach}}, \bibinfo {author} {\bibfnamefont {K.~E.}\ \bibnamefont {McGhee}}, \bibinfo {author} {\bibfnamefont {L.}~\bibnamefont {Ganzer}}, \bibinfo {author} {\bibfnamefont {D.~M.}\ \bibnamefont {Rouse}}, \bibinfo {author} {\bibfnamefont {B.~W.}\ \bibnamefont {Lovett}}, \bibinfo {author} {\bibfnamefont {E.~M.}\ \bibnamefont {Gauger}}, \bibinfo {author} {\bibfnamefont {J.}~\bibnamefont {Keeling}}, \bibinfo {author} {\bibfnamefont {G.}~\bibnamefont {Cerullo}}, \bibinfo {author} {\bibfnamefont {D.~G.}\ \bibnamefont {Lidzey}},\ and\ \bibinfo {author} {\bibfnamefont {T.}~\bibnamefont {Virgili}},\ }\bibfield  {title} {\bibinfo {title} {Superabsorption in an organic microcavity: Toward a quantum battery},\ }\href {https://doi.org/10.1126/sciadv.abk3160} {\bibfield  {journal} {\bibinfo  {journal} {Science advances}\ }\textbf {\bibinfo {volume} {8}},\ \bibinfo {pages} {eabk3160} (\bibinfo {year} {2022})}\BibitemShut {NoStop}%
\bibitem [{\citenamefont {Niu}\ \emph {et~al.}(2024)\citenamefont {Niu}, \citenamefont {Wu}, \citenamefont {Wang}, \citenamefont {Rong},\ and\ \citenamefont {Du}}]{niu2024experimental}%
  \BibitemOpen
  \bibfield  {author} {\bibinfo {author} {\bibfnamefont {Z.}~\bibnamefont {Niu}}, \bibinfo {author} {\bibfnamefont {Y.}~\bibnamefont {Wu}}, \bibinfo {author} {\bibfnamefont {Y.}~\bibnamefont {Wang}}, \bibinfo {author} {\bibfnamefont {X.}~\bibnamefont {Rong}},\ and\ \bibinfo {author} {\bibfnamefont {J.}~\bibnamefont {Du}},\ }\bibfield  {title} {\bibinfo {title} {Experimental investigation of coherent ergotropy in a single spin system},\ }\href {https://doi.org/10.1103/PhysRevLett.133.180401} {\bibfield  {journal} {\bibinfo  {journal} {Physical Review Letters}\ }\textbf {\bibinfo {volume} {133}},\ \bibinfo {pages} {180401} (\bibinfo {year} {2024})}\BibitemShut {NoStop}%
\bibitem [{\citenamefont {Maillette~de Buy~Wenniger}\ \emph {et~al.}(2023)\citenamefont {Maillette~de Buy~Wenniger}, \citenamefont {Thomas}, \citenamefont {Maffei}, \citenamefont {Wein}, \citenamefont {Pont}, \citenamefont {Belabas}, \citenamefont {Prasad}, \citenamefont {Harouri}, \citenamefont {Lema\^{\i}tre}, \citenamefont {Sagnes}, \citenamefont {Somaschi}, \citenamefont {Auff\`eves},\ and\ \citenamefont {Senellart}}]{maillette2023experimental}%
  \BibitemOpen
  \bibfield  {author} {\bibinfo {author} {\bibfnamefont {I.}~\bibnamefont {Maillette~de Buy~Wenniger}}, \bibinfo {author} {\bibfnamefont {S.~E.}\ \bibnamefont {Thomas}}, \bibinfo {author} {\bibfnamefont {M.}~\bibnamefont {Maffei}}, \bibinfo {author} {\bibfnamefont {S.~C.}\ \bibnamefont {Wein}}, \bibinfo {author} {\bibfnamefont {M.}~\bibnamefont {Pont}}, \bibinfo {author} {\bibfnamefont {N.}~\bibnamefont {Belabas}}, \bibinfo {author} {\bibfnamefont {S.}~\bibnamefont {Prasad}}, \bibinfo {author} {\bibfnamefont {A.}~\bibnamefont {Harouri}}, \bibinfo {author} {\bibfnamefont {A.}~\bibnamefont {Lema\^{\i}tre}}, \bibinfo {author} {\bibfnamefont {I.}~\bibnamefont {Sagnes}}, \bibinfo {author} {\bibfnamefont {N.}~\bibnamefont {Somaschi}}, \bibinfo {author} {\bibfnamefont {A.}~\bibnamefont {Auff\`eves}},\ and\ \bibinfo {author} {\bibfnamefont {P.}~\bibnamefont {Senellart}},\ }\bibfield  {title} {\bibinfo {title} {Experimental analysis of energy transfers between a quantum emitter and light fields},\ }\href
  {https://doi.org/10.1103/PhysRevLett.131.260401} {\bibfield  {journal} {\bibinfo  {journal} {Phys. Rev. Lett.}\ }\textbf {\bibinfo {volume} {131}},\ \bibinfo {pages} {260401} (\bibinfo {year} {2023})}\BibitemShut {NoStop}%
\bibitem [{\citenamefont {Hu}\ \emph {et~al.}(2026)\citenamefont {Hu}, \citenamefont {Liu}, \citenamefont {Zhao}, \citenamefont {Zhong}, \citenamefont {Zhou}, \citenamefont {Liu}, \citenamefont {Yuan}, \citenamefont {Lin}, \citenamefont {Xu}, \citenamefont {Hu}, \citenamefont {Xie}, \citenamefont {Liu}, \citenamefont {Zhou}, \citenamefont {Ri}, \citenamefont {Zhang}, \citenamefont {Deng}, \citenamefont {Saguia}, \citenamefont {Linpeng}, \citenamefont {Sarandy}, \citenamefont {Liu}, \citenamefont {Santos}, \citenamefont {Tan},\ and\ \citenamefont {Yu}}]{hu2026quantum}%
  \BibitemOpen
  \bibfield  {author} {\bibinfo {author} {\bibfnamefont {C.-K.}\ \bibnamefont {Hu}}, \bibinfo {author} {\bibfnamefont {C.}~\bibnamefont {Liu}}, \bibinfo {author} {\bibfnamefont {J.}~\bibnamefont {Zhao}}, \bibinfo {author} {\bibfnamefont {L.}~\bibnamefont {Zhong}}, \bibinfo {author} {\bibfnamefont {Y.}~\bibnamefont {Zhou}}, \bibinfo {author} {\bibfnamefont {M.}~\bibnamefont {Liu}}, \bibinfo {author} {\bibfnamefont {H.}~\bibnamefont {Yuan}}, \bibinfo {author} {\bibfnamefont {Y.}~\bibnamefont {Lin}}, \bibinfo {author} {\bibfnamefont {Y.}~\bibnamefont {Xu}}, \bibinfo {author} {\bibfnamefont {G.}~\bibnamefont {Hu}}, \bibinfo {author} {\bibfnamefont {G.}~\bibnamefont {Xie}}, \bibinfo {author} {\bibfnamefont {Z.}~\bibnamefont {Liu}}, \bibinfo {author} {\bibfnamefont {R.}~\bibnamefont {Zhou}}, \bibinfo {author} {\bibfnamefont {Y.}~\bibnamefont {Ri}}, \bibinfo {author} {\bibfnamefont {W.}~\bibnamefont {Zhang}}, \bibinfo {author} {\bibfnamefont {R.}~\bibnamefont {Deng}}, \bibinfo {author} {\bibfnamefont
  {A.}~\bibnamefont {Saguia}}, \bibinfo {author} {\bibfnamefont {X.}~\bibnamefont {Linpeng}}, \bibinfo {author} {\bibfnamefont {M.~S.}\ \bibnamefont {Sarandy}}, \bibinfo {author} {\bibfnamefont {S.}~\bibnamefont {Liu}}, \bibinfo {author} {\bibfnamefont {A.~C.}\ \bibnamefont {Santos}}, \bibinfo {author} {\bibfnamefont {D.}~\bibnamefont {Tan}},\ and\ \bibinfo {author} {\bibfnamefont {D.}~\bibnamefont {Yu}},\ }\bibfield  {title} {\bibinfo {title} {Quantum charging advantage in superconducting solid-state batteries},\ }\href {https://doi.org/10.1103/sp5l-c6m8} {\bibfield  {journal} {\bibinfo  {journal} {Phys. Rev. Lett.}\ }\textbf {\bibinfo {volume} {136}},\ \bibinfo {pages} {060401} (\bibinfo {year} {2026})}\BibitemShut {NoStop}%
\bibitem [{\citenamefont {Baumgratz}\ \emph {et~al.}(2014)\citenamefont {Baumgratz}, \citenamefont {Cramer},\ and\ \citenamefont {Plenio}}]{baumgratz2014quantifying}%
  \BibitemOpen
  \bibfield  {author} {\bibinfo {author} {\bibfnamefont {T.}~\bibnamefont {Baumgratz}}, \bibinfo {author} {\bibfnamefont {M.}~\bibnamefont {Cramer}},\ and\ \bibinfo {author} {\bibfnamefont {M.~B.}\ \bibnamefont {Plenio}},\ }\bibfield  {title} {\bibinfo {title} {Quantifying coherence},\ }\href {https://doi.org/10.1103/PhysRevLett.113.140401} {\bibfield  {journal} {\bibinfo  {journal} {Physical review letters}\ }\textbf {\bibinfo {volume} {113}},\ \bibinfo {pages} {140401} (\bibinfo {year} {2014})}\BibitemShut {NoStop}%
\bibitem [{\citenamefont {Radcliffe}(1971)}]{radcliffe1971some}%
  \BibitemOpen
  \bibfield  {author} {\bibinfo {author} {\bibfnamefont {J.~M.}\ \bibnamefont {Radcliffe}},\ }\bibfield  {title} {\bibinfo {title} {Some properties of coherent spin states},\ }\href {https://doi.org/10.1088/0305-4470/4/3/009} {\bibfield  {journal} {\bibinfo  {journal} {Journal of Physics A: General Physics}\ }\textbf {\bibinfo {volume} {4}},\ \bibinfo {pages} {313} (\bibinfo {year} {1971})}\BibitemShut {NoStop}%
\bibitem [{\citenamefont {Arecchi}\ \emph {et~al.}(1972)\citenamefont {Arecchi}, \citenamefont {Courtens}, \citenamefont {Gilmore},\ and\ \citenamefont {Thomas}}]{arecchi1972atomic}%
  \BibitemOpen
  \bibfield  {author} {\bibinfo {author} {\bibfnamefont {F.~T.}\ \bibnamefont {Arecchi}}, \bibinfo {author} {\bibfnamefont {E.}~\bibnamefont {Courtens}}, \bibinfo {author} {\bibfnamefont {R.}~\bibnamefont {Gilmore}},\ and\ \bibinfo {author} {\bibfnamefont {H.}~\bibnamefont {Thomas}},\ }\bibfield  {title} {\bibinfo {title} {Atomic coherent states in quantum optics},\ }\href {https://doi.org/10.1103/PhysRevA.6.2211} {\bibfield  {journal} {\bibinfo  {journal} {Physical Review A}\ }\textbf {\bibinfo {volume} {6}},\ \bibinfo {pages} {2211} (\bibinfo {year} {1972})}\BibitemShut {NoStop}%
\bibitem [{\citenamefont {Ghosh}\ \emph {et~al.}(2021)\citenamefont {Ghosh}, \citenamefont {Chanda}, \citenamefont {Mal},\ and\ \citenamefont {Sen}}]{ghosh2021fast}%
  \BibitemOpen
  \bibfield  {author} {\bibinfo {author} {\bibfnamefont {S.}~\bibnamefont {Ghosh}}, \bibinfo {author} {\bibfnamefont {T.}~\bibnamefont {Chanda}}, \bibinfo {author} {\bibfnamefont {S.}~\bibnamefont {Mal}},\ and\ \bibinfo {author} {\bibfnamefont {A.}~\bibnamefont {Sen}},\ }\bibfield  {title} {\bibinfo {title} {Fast charging of a quantum battery assisted by noise},\ }\href {https://doi.org/10.1103/PhysRevA.104.032207} {\bibfield  {journal} {\bibinfo  {journal} {Physical Review A}\ }\textbf {\bibinfo {volume} {104}},\ \bibinfo {pages} {032207} (\bibinfo {year} {2021})}\BibitemShut {NoStop}%
\bibitem [{\citenamefont {Hovhannisyan}\ \emph {et~al.}(2013)\citenamefont {Hovhannisyan}, \citenamefont {Perarnau-Llobet}, \citenamefont {Huber},\ and\ \citenamefont {Ac{\'\i}n}}]{hovhannisyan2013entanglement}%
  \BibitemOpen
  \bibfield  {author} {\bibinfo {author} {\bibfnamefont {K.~V.}\ \bibnamefont {Hovhannisyan}}, \bibinfo {author} {\bibfnamefont {M.}~\bibnamefont {Perarnau-Llobet}}, \bibinfo {author} {\bibfnamefont {M.}~\bibnamefont {Huber}},\ and\ \bibinfo {author} {\bibfnamefont {A.}~\bibnamefont {Ac{\'\i}n}},\ }\bibfield  {title} {\bibinfo {title} {Entanglement generation is not necessary for optimal work extraction},\ }\href {https://doi.org/10.1103/PhysRevLett.111.240401} {\bibfield  {journal} {\bibinfo  {journal} {Physical review letters}\ }\textbf {\bibinfo {volume} {111}},\ \bibinfo {pages} {240401} (\bibinfo {year} {2013})}\BibitemShut {NoStop}%
\bibitem [{\citenamefont {Vidal}\ and\ \citenamefont {Werner}(2002)}]{vidal2002computable}%
  \BibitemOpen
  \bibfield  {author} {\bibinfo {author} {\bibfnamefont {G.}~\bibnamefont {Vidal}}\ and\ \bibinfo {author} {\bibfnamefont {R.~F.}\ \bibnamefont {Werner}},\ }\bibfield  {title} {\bibinfo {title} {Computable measure of entanglement},\ }\href {https://doi.org/10.1103/PhysRevA.65.032314} {\bibfield  {journal} {\bibinfo  {journal} {Physical Review A}\ }\textbf {\bibinfo {volume} {65}},\ \bibinfo {pages} {032314} (\bibinfo {year} {2002})}\BibitemShut {NoStop}%
\bibitem [{\citenamefont {Plenio}(2005)}]{plenio2005logarithmic}%
  \BibitemOpen
  \bibfield  {author} {\bibinfo {author} {\bibfnamefont {M.~B.}\ \bibnamefont {Plenio}},\ }\bibfield  {title} {\bibinfo {title} {Logarithmic negativity: a full entanglement monotone that is not convex},\ }\href {https://doi.org/10.1103/PhysRevLett.95.090503} {\bibfield  {journal} {\bibinfo  {journal} {Physical review letters}\ }\textbf {\bibinfo {volume} {95}},\ \bibinfo {pages} {090503} (\bibinfo {year} {2005})}\BibitemShut {NoStop}%
\bibitem [{\citenamefont {Qiu}\ \emph {et~al.}(2023)\citenamefont {Qiu}, \citenamefont {Grimsmo}, \citenamefont {Peng}, \citenamefont {Kannan}, \citenamefont {Lienhard}, \citenamefont {Sung}, \citenamefont {Krantz}, \citenamefont {Bolkhovsky}, \citenamefont {Calusine}, \citenamefont {Kim} \emph {et~al.}}]{qiu2023broadband}%
  \BibitemOpen
  \bibfield  {author} {\bibinfo {author} {\bibfnamefont {J.~Y.}\ \bibnamefont {Qiu}}, \bibinfo {author} {\bibfnamefont {A.}~\bibnamefont {Grimsmo}}, \bibinfo {author} {\bibfnamefont {K.}~\bibnamefont {Peng}}, \bibinfo {author} {\bibfnamefont {B.}~\bibnamefont {Kannan}}, \bibinfo {author} {\bibfnamefont {B.}~\bibnamefont {Lienhard}}, \bibinfo {author} {\bibfnamefont {Y.}~\bibnamefont {Sung}}, \bibinfo {author} {\bibfnamefont {P.}~\bibnamefont {Krantz}}, \bibinfo {author} {\bibfnamefont {V.}~\bibnamefont {Bolkhovsky}}, \bibinfo {author} {\bibfnamefont {G.}~\bibnamefont {Calusine}}, \bibinfo {author} {\bibfnamefont {D.}~\bibnamefont {Kim}}, \emph {et~al.},\ }\bibfield  {title} {\bibinfo {title} {Broadband squeezed microwaves and amplification with a josephson travelling-wave parametric amplifier},\ }\href {https://doi.org/10.1038/s41567-022-01929-w} {\bibfield  {journal} {\bibinfo  {journal} {Nature Physics}\ }\textbf {\bibinfo {volume} {19}},\ \bibinfo {pages} {706} (\bibinfo {year} {2023})}\BibitemShut
  {NoStop}%
\bibitem [{\citenamefont {Bienfait}\ \emph {et~al.}(2016)\citenamefont {Bienfait}, \citenamefont {Pla}, \citenamefont {Kubo}, \citenamefont {Zhou}, \citenamefont {Stern}, \citenamefont {Lo}, \citenamefont {Weis}, \citenamefont {Schenkel}, \citenamefont {Vion}, \citenamefont {Esteve} \emph {et~al.}}]{bienfait2016controlling}%
  \BibitemOpen
  \bibfield  {author} {\bibinfo {author} {\bibfnamefont {A.}~\bibnamefont {Bienfait}}, \bibinfo {author} {\bibfnamefont {J.}~\bibnamefont {Pla}}, \bibinfo {author} {\bibfnamefont {Y.}~\bibnamefont {Kubo}}, \bibinfo {author} {\bibfnamefont {X.}~\bibnamefont {Zhou}}, \bibinfo {author} {\bibfnamefont {M.}~\bibnamefont {Stern}}, \bibinfo {author} {\bibfnamefont {C.}~\bibnamefont {Lo}}, \bibinfo {author} {\bibfnamefont {C.}~\bibnamefont {Weis}}, \bibinfo {author} {\bibfnamefont {T.}~\bibnamefont {Schenkel}}, \bibinfo {author} {\bibfnamefont {D.}~\bibnamefont {Vion}}, \bibinfo {author} {\bibfnamefont {D.}~\bibnamefont {Esteve}}, \emph {et~al.},\ }\bibfield  {title} {\bibinfo {title} {Controlling spin relaxation with a cavity},\ }\href {https://doi.org/10.1038/nature16944} {\bibfield  {journal} {\bibinfo  {journal} {Nature}\ }\textbf {\bibinfo {volume} {531}},\ \bibinfo {pages} {74} (\bibinfo {year} {2016})}\BibitemShut {NoStop}%
\bibitem [{\citenamefont {Aragone}\ \emph {et~al.}(1974)\citenamefont {Aragone}, \citenamefont {Guerri}, \citenamefont {Salamo},\ and\ \citenamefont {Tani}}]{aragone1974intelligent}%
  \BibitemOpen
  \bibfield  {author} {\bibinfo {author} {\bibfnamefont {C.}~\bibnamefont {Aragone}}, \bibinfo {author} {\bibfnamefont {G.}~\bibnamefont {Guerri}}, \bibinfo {author} {\bibfnamefont {S.}~\bibnamefont {Salamo}},\ and\ \bibinfo {author} {\bibfnamefont {J.}~\bibnamefont {Tani}},\ }\bibfield  {title} {\bibinfo {title} {Intelligent spin states},\ }\href {https://doi.org/10.1088/0305-4470/7/15/001} {\bibfield  {journal} {\bibinfo  {journal} {Journal of Physics A: Mathematical, Nuclear and General}\ }\textbf {\bibinfo {volume} {7}},\ \bibinfo {pages} {L149} (\bibinfo {year} {1974})}\BibitemShut {NoStop}%
\bibitem [{\citenamefont {Rashid}(1978)}]{rashid1978intelligent}%
  \BibitemOpen
  \bibfield  {author} {\bibinfo {author} {\bibfnamefont {M.}~\bibnamefont {Rashid}},\ }\bibfield  {title} {\bibinfo {title} {The intelligent states. i. group-theoretic study and the computation of matrix elements},\ }\href {https://doi.org/10.1063/1.523840} {\bibfield  {journal} {\bibinfo  {journal} {Journal of Mathematical Physics}\ }\textbf {\bibinfo {volume} {19}},\ \bibinfo {pages} {1391} (\bibinfo {year} {1978})}\BibitemShut {NoStop}%
\end{thebibliography}%

\end{document}